\documentclass[11pt]{article}
\input{epsf.sty}
\usepackage{hyperref}
\usepackage{amsmath, amssymb} 
\usepackage[pdftex]{graphicx}
\usepackage{subfig}
\usepackage{cite}
\usepackage{appendix}

\newtheorem {thm}{Theorem}[section]
\newtheorem {prop}[thm]{Proposition}
\newtheorem {lem}[thm]{Lemma}
\newtheorem {cor}[thm]{Corollary}
\newtheorem {defn}[thm]{Definition}

\def\Cox{\hfill \Box}

\def\R{{\mathbb R}}
\def\P{{\mathbb P}}

\def\E{{\mathbb E}}
\def\HH{{\cal H}}

\def\0{{\bf 0}}

\def\a{\alpha}
\def\b{\beta}

\def\e{\varepsilon}

\def\phi{\varphi}
\def\g{\gamma}
\def\l{\lambda}

\def\s{\sigma}

\def\D{\Delta}
\def\L{\Lambda}

\def\T{\T}

\def\LL{{\cal L}}

\def\Const{\text{Const}\,}

\begin{document}

\title{Low-temperature dynamics of the Curie-Weiss Model: \\ 
Periodic orbits, multiple histories, and  loss of Gibbsianness} 

\author{
Victor Ermolaev
\footnote{
Rijksuniversiteit Groningen, 
Faculteit Wiskunde en Natuurwetenschappen, 
Nijenborgh 9,
9747 AC Groningen,
The Netherlands,
\texttt{v.n.ermolaev@rug.nl}}~~and 
Christof K\"ulske
\footnote{
Ruhr-Universit\"at Bochum, 
Fakult\"at f\"ur Mathematik, 
Universit\"atsstrasse 150,
44780 Bochum, 
Germany,
\texttt{Christof.Kuelske@rub.de}
}}

\maketitle
%ABSTRACT
\begin{abstract}
We consider the Curie-Weiss model at initial temperature 
$0<\b^{-1}\leq \infty $ in vanishing 
external field evolving under a Glauber spin-flip dynamics with temperature $0<{\b'}^{-1}\leq \infty$.
We study the limiting conditional probabilities and their continuity properties and discuss their set of points of discontinuity (bad points).
We provide a complete analysis of the transition between Gibbsian and non-Gibbsian behavior as a function of time, 
extending earlier work for the case of independent spin-flip dynamics.  

For initial temperature $\b^{-1}>1$ we prove that the time-evolved measure 
stays Gibbs forever, for any (possibly low) temperature of the dynamics. 

In the regime of heating to low-temperatures from even lower temperatures,  
$0<\b^{-1}<\min\{{\b'}^{-1},1\}$ we prove that the time-evolved measure 
is Gibbs initially and becomes non-Gibbs after a sharp transition time. 
We find this regime is further divided into a region where only symmetric bad configurations 
exist, and a region where this symmetry is broken.

In the regime of further cooling from low-temperatures, ${\b'}^{-1}<\b^{-1}<1$ 
there is always symmetry-breaking in the set of bad configurations. These bad configurations are 
created by a new mechanism which is related to the occurrence of periodic 
orbits for the vector field which describes the dynamics of Euler-Lagrange 
equations for the path large deviation functional for the order parameter. 

To our knowledge this is the first example of the rigorous study of non-Gibbsian phenomena related to cooling, albeit 
in a mean-field setup.  
\end{abstract}

\bigskip 


%INTRODUCTION
\section{Introduction} \label{sect:intro}

Non-Gibbsian measures are known to appear in many circumstances. Historically 
they were observed first in the context of position-space renormalization group transformation
and termed as so-called RG pathologies \cite{griffiths1979}.  Later more and more examples 
were discovered \cite{EnFerSok,Pr04,fernandezLesHouches, LeNyEnsanios,vanEdenHRedFer} which showed that the application of many maps applied to an 
infinite-volume Gibbs measure may result in similar "pathologies", meaning that the image measure
is not a Gibbs measure anymore. When such a phenomenon appears it means that conditional 
probabilities of the image system will acquire long-range dependencies, at least for some non-removable 
configurations. 
Particularly interesting examples of infinite-volume transformations are coming from the study of dynamics \cite{vanEdenHRedFer,KuLeNy, ReRoRu10,KuOp08,KuLnRe04}.
The first prototypical result in that direction is due to van Enter, Fernandez, den Hollander, Redig who considered 
an infinite-temperature (or high temperature) Glauber dynamics starting from an initial low-temperature 
Ising model on the two- or more-dimensional integer lattice. In particular they proved that a low-temperature 
initial measure in vanishing external magnetic field becomes non-Gibbs after sufficiently large times and 
stays non-Gibbs forever. This has to be contrasted with the simple fact that for independent dynamics, viewed on local observables, 
the time-evolved measure converges exponentially fast in time to the symmetric product measure. 
In fact such a phenomenon is possible since the Gibbs property (continuity property of conditional probabilities 
of the system) is to be tested in arbitrarily large volumes. 
Later more investigations for time-evolutions were performed. The general picture is that 
for very general dynamics and very general initial measures the time-evolved measures 
are again Gibbsian, for a sufficiently small time-interval \cite{KuOp,LeNyEnsanios,LeNyRe,vanEdenHRedFer,EnRu}.  
Long times however, even for simple dynamics offer the possibility for the emergence of non-Gibbsian measures. 
The discontinuities in the conditional probabilities which are responsible are produced by hidden phase 
transitions which pop up as a result of the conditioning procedure. Depending on the specific 
nature of the system there may be many mechanisms of such singularities \cite{KuLeNy,vanEdenHRedFer}. In this context 
continuous spin models are particularly interesting \cite{KuRe, EnKuOpRu, EnRu}.  

While it is surprising that even the physically simple transformation of heating produces non-Gibbsian behavior
it would even be more interesting to say something about {\em cooling dynamics}. 
More generally one would like to study a Gibbs measure $\mu_0$ for an initial Hamiltonian $H$ which is subjected 
to a Glauber dynamics for another Hamiltonian $\bar H$,  which gives rise to a trajectory $\mu_t$ 
where $t$ denotes time. Glauber dynamics at low temeratures describes fast cooling or ``quenching''. 
The question is to understand the behavior of $\mu_t$, and in particular for which times it will be Gibbs. 
Since this is as yet too difficult on the lattice, we develop our results in mean-field. 
A mean-field system of Ising-type is called non-Gibbs if the single-site conditional probabilities
depend in a discontinuous way on the magnetization of the conditioning spins \cite{HaKu, Ku, KuOp,EnKu}. 
Investigations for mean-field models tend to reproduce the lattice results in many situations \cite{KuOp, SaWre} 
but often lead to an explicit knowledge of the parameter regions where Gibbsianness and non-Gibbsianness 
occur. Such an analysis has been performed for the Curie-Weiss model subjected 
to an independent spin-flip dynamics in \cite{KuLeNy}. 
It was proved there that for initial high temperatures $\b^{-1}\geq 1$ the time-evolved 
measure is Gibbs forever, while for $\b^{-1}<1$ there exists a sharp transition-time 
separating a Gibbsian from a non-Gibbsian regime. 
In the course of the analysis of that paper, also the phenomenon of 
symmetry-breaking in the set of bad configurations was observed which happens for the smaller range of initial temperatures below 
$\frac{2}{3}$. In the present paper we build on that analysis but are able to extend the results to dependent spin-flips 
according to a Glauber dynamics with an arbitrary other temperature ${\b'}^{-1}$.

To understand  discontinuous behavior of conditional probabilities for the time-evolved model at fixed time $t$ 
one needs to look at the model resulting from the initial measure at time $s=0$ under application of the dynamics 
in the space-time region for times $s$ between $0$ and $t$. 
The hidden phase transitions responsible for the non-Gibbsian behavior 
occur if there is a sensitive dependence of the model at time $s=0$ obtained from constraining the space-time 
measure to certain configurations at time $s=t$.  If a small variation of such a constraining configuration leads to a jump in the constrained initial measure it will (generically) 
be a bad configuration for the conditional probabilities of the system at time $t$. 
{\em Small variation} means in the lattice case a perturbation in an annulus far away 
from the origin. {\em Small variation} means in the mean-field case a small variation of the magnetization as a real number.
In the independent spin-flip lattice example of \cite{EnFerSok} the chessboard configuration was a bad one, 
correspondingly in the independent spin-flip mean-field case of \cite{KuLeNy} the configurations with neutral magnetization equal to zero 
were bad ones for large enough times. Moreover, configurations with non-zero magnetization also appeared as points of discontinuity for 
the limiting conditional probabilities, in a particular bounded region of the parameter space of initial temperature and time. 
This phenomenon was called biased non-Gibbsianness in \cite{KuLeNy}. 
The complete analysis for the mean-field independent spin-flip situation was possible since the constrained 
system on the first layer could be understood on the level of the magnetization. The relevant quantities 
could be computed in terms of the rate-function for a standard quenched disordered model, namely the 
Curie-Weiss random-field Ising model with possibly non-symmetric random-field distribution of the quenched disorder. 

To deal with dependent-dynamics case a different route has to be taken since the dependence of the initial system 
on the conditioning is more intricate. As we will see, we will need to invoke the path large deviation principle 
for the dynamics with temperature ${\b'}^{-1}$ on the level of magnetizations. 
We will then have to minimize a cost functional of paths of magnetizations 
which is composed of the rate function along the path and an initial ``punishment'' term, which depends both on 
the initial Hamiltonian $H$ and the dynamical Hamiltonian $\bar H$, evaluated at the unknown initial point of
the trajectory. The solution of the problem gives a surprising connection between path properties of the corresponding (integrable) 
dynamical system and Gibbs properties of a model of statistical mechanics. 
As a result we are providing a full description of the regions of Gibbsian and non-Gibbsian behavior as a function of time, 
initial temperature, and dynamical temperature. As a special case the previous results for infinite-temperature dynamics 
are reproduced (adding some geometrical insight about the behavior of typical paths). 
Furthermore the solution reveals a new mechanism for the appearance of bad configurations 
in the region of cooling from low temperatures with even lower temperatures. 
These are related to periodic motion in the dynamical system. 

The present paper is to our knowledge the first one where Gibbs properties 
of a model subjected to a low-temperature dynamics are investigated and it will be challenging 
to see which parts of the behavior are occurring  on the lattice. After the completion of our work 
we learnt about the preprint \cite{EnFerdenHRed2010}
where a large-deviation approach was proposed to understand dynamical transitions in the Gibbs 
properties for lattice systems, too. While there is a beautiful formalism available for path 
large deviations of empirical measures of lattice systems on an abstract level, explicit results are very hard, 
which underlines also the use of our present paper, and the necessity of future research.

Moreover the questions and methods used should have interest also in models 
of population dynamics. In such models a population of $N$ individuals, each individual 
carrying genes from a finite alphabet of possible types, performs a stochastic dynamics which 
can be described on the level of empirical distributions. Starting the dynamics from 
a known initial measure corresponds to an a-priori belief (prior distribution) over the distribution of types. 
Conditioning to a final configuration $m'$ at time $t$ corresponds to measuring 
the distribution of types. The occurrence of multiple histories leading to the same $m'$
(which is responsible for non-Gibbsianness in the spin-model) 
has the interesting interpretation of a non-unique best estimator for the 
path explaining the present mix of genes. 
\bigskip
\bigskip

\subsection{The model at time t=0}

We start at time $s=0$ with the Curie-Weiss Ising model 
in zero magnetic field at inverse temperature $\b$ whose 
finite-volume Gibbs measures 
on spin-configurations $\sigma_{[1,N]}=(\s_i)_{i=1,\dots,N}\in\{-1,1\}^N$
are given by 
\begin{equation}\label{eq:inmeas}
\mu_{\b,N}(\s_{[1, N]})
=\frac{\exp\Bigl(\frac{\b}{2 N}\bigl(\sum_{i=1}^N \s_i\bigr)^2  
\Bigr)}{Z_{\b,N}}
\end{equation}
where the normalization factor $Z_{\b, N}$ is the standard
partition function. 
This model shows a phase transition at the critical temperature 
$\b^{-1}=1$ in the limit where $N\rightarrow\infty$.

\subsection{The dynamics}
Given a configuration $\sigma_{[1,N]}=(\s_i)_{i=1,\dots,N}\in\{-1,1\}^N$ of spins,
we set Glauber dynamics on the level of the spins in such a way that it has 
the Curie-Weiss distribution with a (possibly) different temperature ${\b'}^{-1}$ as 
a reversible measure. We will call ${\b'}^{-1}$ the dynamical temperature.
The generator of the system with $N$ spins is given by 
\begin{equation}\label{eq:linGen}
\begin{split}
L_N \Phi(\s_{[1,N]})=\sum_{i=1}^N c\bigl(\s_i,\frac{1}{N}\sum_{j:j\neq i}\s_j\bigr)
\Bigl(\Phi(\s^{i}_{[1,N]})-\Phi(\s_{[1,N]})\Bigr)\end{split}
\end{equation}
where $\s^{i}_{[1,N]}$ denotes the configuration that is flipped at the site $i$
\begin{equation}
\begin{split}
\left(\sigma^i_{[1,N]}\right)_j = 
\left\{
		\begin{array}{rl}
		-\left(\sigma^i_{[1,N]}\right)_i,& j=i\\
		\left(\sigma^i_{[1,N]}\right)_j,& j\neq i
		\end{array}
\right.
\end{split}
\end{equation}
where we choose the rates to be 
\begin{equation}
\begin{split}
c(\mp ,m)=\frac{e^{\pm \b' m}}{\cosh(\b' m)-m\sinh(\b' m)}  
\end{split}
\end{equation}

For fixed finite $N$, we denote the corresponding time-evolved measure on $\{-1,1\}^N$ at time $t$, 
started from the equilibrium measure $\mu_{\b,N}$,  by the symbol $\mu_{\b,\b',t;N}$.
It is clear that, for fixed $N$, the time-evolved 
measure $\mu_{\b,\b',t;N}$ tends to the invariant measure under the dynamics, when $t\uparrow\infty$. 

\subsection{The notion of Gibbsianness for mean-field models}

For a single-site spin 
$\s_1 \in \left\{-1,+1\right\}$ and a magnetization value 
for a system of size $N-1$, that is $
\widehat m \in \{-1,-1 + \frac{2}{N-1},\dots, 1-\frac{2}{N-1},1\}$ we 
consider the single-site conditional probabilities of the time-evolved measure in the volume $N$ given by 
\begin{equation}
\label{eq:cond-prob1}
\g_{\b,\b',t,N}(\s_1| \widehat{m})
:=\mu_{\b,\b',t,N}(\s_1| \s_{[2,N]}),
\end{equation}
where $\s_{[2,N]}$ is any spin-configuration 
such that  $\widehat{m}=\frac{1}{N-1}\sum_{j=2}^N \s_j$.
By permutation invariance the right-hand side of \eqref{eq:cond-prob1} does not depend on the choice of $\s_{[2,N]}$.

\begin{defn}\label{def:goodConf}
Let $\b, \b', t$ be given. A point $\widehat{m} \in (-1,1)$ is said to be {\em good} for the time-evolved 
mean-field model if and only if 
\begin{enumerate}
\item 
	There exists a neighborhood of $\widehat{m}$ such that, for all $\a$ in this neighborhood
	the following holds. For all sequences $\a_N \in \{-1,-1 + \frac{2}{N-1},\dots, 1-\frac{2}{N-1},1\}$
	with the property $\lim_{N\uparrow\infty}\a_N=\a$ the limit 
	\begin{equation}
	\g_{\b,\b',t}(\s_1 |\a)=\lim_{N\uparrow \infty}\g_{\b,\b',t,N}(\s_1| \a_N)
	\end{equation}
	exists and is independent of the choice of the sequence. 

\item
	 The function $\a \mapsto \g_{\b,\b',t}(\s_1 |\a)$ is continuous at $\a=\widehat{m}$. 
\end{enumerate}
\end{defn}

\begin{defn}
\label{def:mean-field-gibbs}
The time-evolved mean-field model with parameters $\b,\b',t$ is called {\em Gibbs} iff it has no bad points.
\end{defn}

This definition has extensions to arbitrary local state spaces beyond finite types 
(see \cite{HaKu,KuOp}) where empirical magnetizations have to be replaced by empirical distributions in the definition. 
\bigskip 

\subsection{Main Theorem} 

We are now in the position to give our main result. 

\begin{thm} Consider the time-evolved Curie-Weiss model with initial and dynamical temperatures ${\b}^{-1},{\b'}^{-1}$. 

Then the following holds. 

\begin{enumerate}
\item Initial high temperature, any temperature of the dynamics. \\
If $\b^{-1}\geq 1$ then the time-evolved model is Gibbs for all $t\geq 0$. 

\item Heating from an initial low-temperature, with a either high-temperature or a low-temperature dynamics.\\
For any $\b'$ there exists a value $\b^{-1}_{\text{SB}}(\b')<{\b'}^{-1}$ (which is explicitly computable, see below) 
such that the following is true. 

Assume that $0<\b^{-1}<\min\{{\b'}^{-1},1\}$. 

\begin{enumerate}
\item If $\b^{-1}_{\text{SB}}(\b')\leq\b^{-1}$ then 

\begin{itemize}
\item for all $0 \leq t \leq t_{nGS}(\b,\b'):=\frac{\ln\frac{\b'-\b}{1-\b}}{4(1-\b')}$ the time-evolved model is Gibbs. 

\item for all $t > t_{nGS}(\b,\b')$ the model is not Gibbs and the time-evolved conditional probabilities are discontinuous at 
$\widehat m = 0$ and continuous at any $\widehat m \neq 0$. 

\end{itemize}

\item 
If $0<\b^{-1}<\b^{-1}_{\text{SB}}(\b')$ there exist sharp values $0<t_0(\b,\b')<t_1(\b,\b')<\infty$ 
such that 

\begin{itemize}
\item for all $0 \leq t \leq t_{0}(\b,\b')$ the time-evolved model is Gibbs,  

\item for all $t_0(\b,\b')<t<t_1(\b,\b')$ there exists $\widehat{m}_c=\widehat{m}_c(\b,\b';t)\in (0,1)$ 
such that the limiting conditional probabilities are discontinuous at the points $\pm \widehat{m_c}$, 
and continuous otherwise, 

\item  for all $t > t_1(\b,\b')$ the limiting conditional probabilities are discontinuous at $\widehat{m}=0$
and continuous at any $\widehat{m} \neq 0$.

\end{itemize}

\end{enumerate}

\item  Cooling from initial low temperature.
For ${\b'}^{-1}<\b^{-1}<1$ there exists a time-threshold $t_{\hbox{per}}(\b,\b')$ such that,

\begin{itemize}
\item for all $0 \leq t \leq t_{\hbox{per}}(\b,\b')$ the time-evolved model is Gibbs. 

\item for all $t > t_{\hbox{per}}(\b,\b')$ the model is not Gibbs and the time-evolved conditional probabilities are discontinuous at non-zero configurations $\widehat{m}_c$ (and continuous at $\widehat{m}=0$). 

\end{itemize}
\end{enumerate}
\end{thm}
\begin{figure}[htb]
\centering
\includegraphics[height=7.5cm]{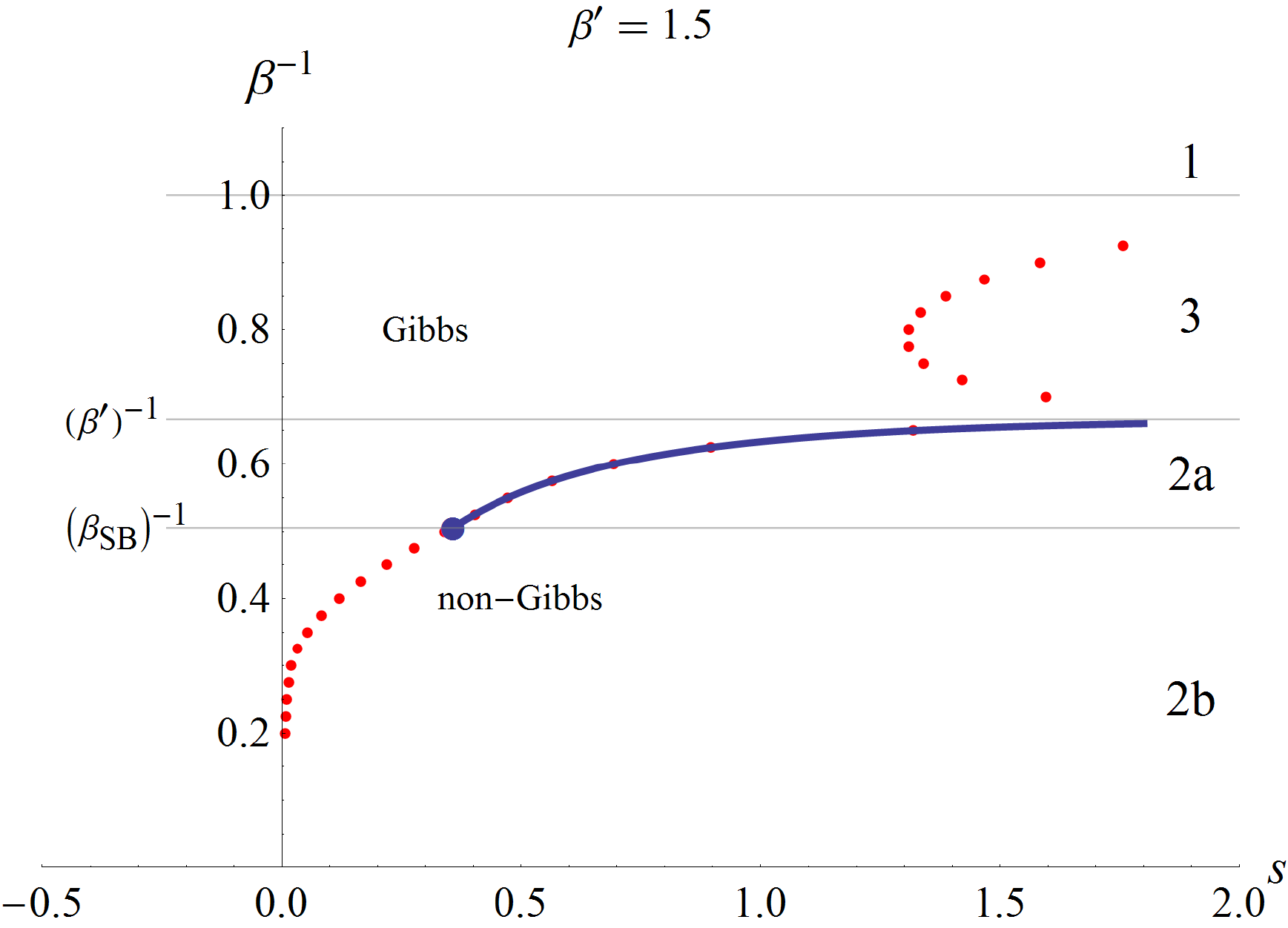}
\caption{Division between Gibbs and non-Gibbs area for low-temperature dynamics, the thick 
curve is obtained by computation, the dots are given by numerics}
\label{fig:gng}% gibbs non gibbs
\end{figure}

Note that for high-temperature dynamics ${\b'}^{-1}>1$ the region 3 of initial temperatures 
in Figure \ref{fig:gng}  is empty.
Part 2 of the theorem generalizes the structure which we already know from the independent spin-flip dynamics ${\b'}=0$ (see \cite{KuLeNy})  
which is contained as a special case. 
This means that a symmetric (w.r.t. starting measure) bad point $m_0=0$
will appear after a sharp transition time if the initial temperature is not too low (see Subregion 2a). 
For lower temperatures (in Subregion 2b) symmetry-breaking in the set of bad configurations
for the time-evolved measure appears in an intermediate time-interval: At the beginning 
of this interval a symmetric pair of bad configuration appears which merges at the 
end of the time interval. 

It is remarkable that the picture we observe in Region 2 is similar to the independent spin-flip case. 
This is even true for low temperatures ${\b'}^{-1}<1$ of the dynamics. 
As we will see, we can moreover compute the symmetry-breaking 
inverse temperature $\b_{SB}$ in terms of $\b'$ as the largest solution of the following cubic equation
\begin{equation} 
 \begin{split}
	4 \b^3_{\text{SB}} + 12 \b_{\text{SB}} \b' - 6 \b_{\text{SB}}^2 (1 + \b') - \b' (3 + 3\b' - {\b'}^2)=0
\end{split}
\end{equation}
In the independent spin-flip case ${\b'}=0$ we get exactly $\b^{-1}=\frac23$, which 
was already found in \cite{KuLeNy}.
We will also give an explicit expression of the critical time in region 2a, for all $\b'$.   

In region 3 of cooling from an already low initial temperature we observe an 
entirely new mechanism for the production of non-Gibbsian points. These are related
to periodic orbits of the flow of the $\b'$-dependent vector field
which is created by the Euler-Lagrange equations obtained from the path-large-deviation principle for
the given dynamics.

\subsection{Strategy of proof and phase-space picture}

To derive an expression for the time-evolved kernel $\g_{\b,\b',t}$ it turns
out that we need to look at path large deviations of the evolving empirical magnetization,
on a fixed time-interval $[0,t]$ with $N$ as a large parameter, conditioned to end in a
fixed magnetization $m' \in (-1,1)$.
The path large deviation functional consists of two parts and can be viewed as a Lagrangian
on the space of paths of magnetization on $[0,t]$.
The first one is an integral over the time interval 
of a Lagrangian density depending on $\b'$ as a parameter,
and also on the magnetization variable and its time-derivative. 
Since the dynamics is started from an initial measure, the rate-functional in the LDP will contain
also a second $\b$-dependent term ``punishing'' the choice of the (unknown) initial-condition.
The solution of the corresponding path minimization problem 
will therefore depend on a balance between both terms. Such solution (or solutions, in case of multiple minima) will correspond to a most probable history path(s). Non-uniqueness of the solution makes possible
%This balance is responsible for the possibility of 
a jump of the most probable history curve which ends at a prescribed final condition $m'$ when one varies 
around particular values of $m'$. These particular values will become  
discontinuity points of $\g_{\b,\b',t}(m')$. 
The problem of finding the most probable conditioned history path 
carries over analytically to the study of the evolution of 
a curve describing the allowed initial conditions for the magnetization and its velocity 
(depending on $\b,\b'$) under the flow of the Euler-Lagrange equations (depending on $\b'$). 
Multiple histories show in this framework as multiple projections of the time-evolved 
curve in phase-space to the $m$-axis, and this will allow us to derive geometric insight 
as well as analytical and numerical results. As a warning we point out that the notion of ``Hamiltonian''
will always refer to a spin-Hamiltonian, not the Legendre transform of the discussed Lagrangian.

The outline of the paper is as follows.
Section 2 will be devoted to the derivation of 
the path large deviation principle,
as well as the constrained large deviation principle involving the initial Hamiltonian and its 
relation to the time-evolved conditional probabilities. 
In Section 3 we discuss the solution of the variational problem in terms of the Euler-Lagrange 
equations giving rise to a time-evolved curve of allowed initial configurations.  
Section 4 provides more visual intuition for the system's behavior based on numerics.  

\subsection{Acknowledgements}
We thank Aernout van Enter, Roberto Fernandez, Frank den Hollander, Frank Redig, and 
Evgeny Verbitskiy for stimulating discussions during the Groningen Nature-Nurture meetings.  
C.K. thanks Anton Bovier and Amir Dembo for enlightening comments on path large deviations. 

\bigskip
\bigskip

\section{Path large deviation principle and limiting conditional probabilities} 

Before we start discussing a number of large-deviantions results it is appropriate to
rewrite the finite-volume Gibbs measure \eqref{eq:inmeas} on spin-configurations $\s_{[1,N]}$ as follows
\begin{equation}
\mu_{\b,N}(\s_{[1, N]}) = \frac{\exp\left(-N H(m_N)\right)}{Z_{\b,N}},
\end{equation}
where $m_N:\s_{[1,N]}\mapsto \frac{1}{N}\sum_{i=1}^N\s_i$ is 
the function which sends a spin configuration to its empirical mean and $H(x):=-\frac{\b x^2}{2}$ is 
the spin-Hamiltonian 
of the system.

In this section we will provide an expression for the limiting conditional probabilities. 
This involves the large-$N$ asymptotics for the paths of the empirical magnetization. 
Note first that, by permutation invariance, the continuous time process that is induced on the empirical magnetization is again 
a Markov chain. 

Namely,  
suppose that $F:\{-1,-1+\frac{2}{N}, \dots, 1-\frac{2}{N},1 \}\rightarrow \R$ is a function on the 
possible magnetization values at size $N$ and $m_N$ is an empirical mean,
% Namely,  
% suppose that $F:\{-1,-1+\frac{2}{N}, \dots, 1-\frac{2}{N},1 \}\rightarrow \R$ is a function on the 
% possible magnetization values at size $N$ and $m_N:\s_{[1,N]}\mapsto \frac{1}{N}\sum_{i=1}^N\s_i$ is 
% the function which sends a spin configuration to its empirical mean,
then we have $L_{\b',N} (F\circ m_N)=(\hat L_{\b',N} F)\circ m_N$ with 
\begin{equation}
\label{eq:lin-genm}
\begin{split}
&\hat L_{\b',N} F(m)=\Bigl( \frac{N}{2} +\frac{N}{2}m \Bigr)
c\bigl(+,m -\frac{1}{N}\bigr)
\Biggl(F\Bigl(m-\frac{2}{N}
\Bigr)- F\Bigl(m\Bigr)\Biggr)\cr
&+\Bigl( \frac{N}{2} -\frac{N}{2} m \Bigr)
c\bigl(-,m+\frac{1}{N}\bigr)
\Biggl(F\Bigl(m +\frac{2}{N}
\Bigr)- F\Bigl(m\Bigr)\Biggr)
\end{split}
\end{equation} 
How do typical paths for the unconstrained dynamics look for large $N$? 
Evaluating  $\frac{d}{dt}\E^{m_0}(F(m_t))=\E^{m_0}( (\hat L_{\b',N} F)(m_t))$ for the observable $F(m)=m$ 
for the expected value of the process started at $m_0$ we have the identity 
\begin{equation}
\begin{split}
\frac{d}{dt}\E_N^{m_0}m_t = \E_N^{m_0}\left[(1-m_t)c(-,m_t+\frac{1}{N})-(1+m_t)c(+,m_t-\frac{1}{N})\right]
\end{split}
\end{equation}
Taking the limit $N\rightarrow\infty$ %under the assumption that the distribution of the paths of 
the magnetization concentrates on a deterministic path $t\mapsto m(t)$ 
%(as we will discuss in more detail in a moment) 
which solves the ODE $\dot m = (1-m)c(-,m)-(1+m)c(+,m)$ or 
\begin{equation}\label{eq:deterministicR} \dot m = 2 \frac{ \sinh(\b' m)- m \cosh(\b' m) }{ \cosh(\b' m)-m \sinh(\b' m)}
\end{equation}
which has the largest solution of the mean-field equation $m=\tanh(\b' m)$ as a stable solution. 
In the case $\b' = 0$ the equation reduces to the linear equation 
$\dot m(t)=-2m(t)$ which describes the relaxation of the magnetization to zero 
under the unconstrained infinite-temperature dynamics. 

Next we need to discuss a number of large deviation results which are needed 
to compute the limiting conditional probabilities. 
We begin as the first ingredient 
with the statement of the path large deviation principle 
for the dynamics with inverse temperature $\b'$. 

\begin{thm}\label{thm:formJ}
Denote by $P_{\b', N}$ the law of the paths 
$(z_{N}(s))_{s\in [0,t]}$ of the magnetization for the continuous-time 
Markov-chain with generator $L_{\b',N}$. 

Then the measures $P_{\b',N}$ satisfy 
% in $L_{\infty}[0,t]$ 
a large deviation principle 
with rate $N$  and rate function given by the Lagrange functional  
$$\phi\mapsto J_{\b'}(\phi)=\int_{0}^t j_{\b'}(\phi(s),\dot \phi(s))ds$$ 
with Lagrange density $j_{\b'}(m,v)$ given by 

\begin{equation}\label{eq:fullJ}
\begin{split}
j_{\b'}&(m,v)=\frac{1}{2}\Biggl\{ 2-\sqrt{\frac{e^{4 \b' m} (-1+m)^2v^2
+(1+m)^2 v^2-2 e^{2 \b' m} \left(-1+m^2\right) \left(8+v^2\right)}{\left(1-e^{2\b' m} (-1+m)+m\right)^2}}\cr
&+v \log\left[\frac{e^{-2\b' m}\left(-1+e^{2\b' m} (-1+m)-m\right)}{4(-1+m)}\right]\cr
&+v \log\left[v+\sqrt{\frac{e^{4\b' m} (-1+m)^2 v^2+(1+m)^2 v^2-2 e^{2\b' m}
\left(-1+m^2\right) \left(8+v^2\right)}{\left(1-e^{2\b' m}
(-1+m)+m\right)^2}}\right]\Biggr\}
\end{split}
\end{equation}
\end{thm}

For the special important case of non-interacting dynamics $\b'=0$ we write
\begin{equation}\label{eq:simpleJ}
\begin{split}
j(m,v):=j_{0}(m,v)=\frac{1}{2} \left(2-\sqrt{4-4 m^2+v^2}+v 
\log\left[\frac{v+\sqrt{4-4 m^2+v^2}}{2-2 m}\right]\right)
\end{split}
\end{equation}
The proof of Theorem \ref{thm:formJ} will be sketched in the Appendix. 
%postponed to the end of this Section. 
This large deviation principle allows us to compute the large deviation 
asymptotics of finding the path of the magnetization jump process at finite $N$
close to a given path $\phi(t)$. 

It allows us to compute the large deviation asymptotics of the probability to 
go from an initial configuration $m$ to a final condition $m'$ in time $t$ 
by computing the value of the rate function in the minimizing path from $m$ to $m'$. 
The minimizing path is found by solving the Euler-Lagrage equations 
with initial condition $m$ and final condition $m'$. \bigskip 

The second and more elementary ingredient we need is the static large deviation principle for 
the magnetization in the initial measure, the Curie-Weiss measure with inverse
temperature $\b$. It reads as follows. 
\begin{prop}\label{thm:formJ1}
The distribution of magnetization $m=\frac{1}{N}\sum_{i=1}^N \s_i$ 
w.r.t. the Curie-Weiss measure at inverse temperature $\b$ obeys 
a large deviation principle 
with rate $N$ and rate function given by 
%$-\frac{\b m^2}{2}+I(m)$ where
$H(m)+I(m)$ where  
\begin{equation}\label{eq:II}
\begin{split}
I(m)=\frac{1+m}{2}\log(1+m)+\frac{1-m}{2}\log(1-m)
\end{split}
\end{equation}
is the rate function for the symmetric Bernoulli distribution. 
\end{prop}

The Proposition is well known in the theory of mean-field systems. 
It follows from Varadhan's Lemma which states the following: 
Suppose a probability distribution satisfies a LDP principle with a known rate 
function and rate $N$ and suppose we consider the probability distribution 
with density $C e^{- N H(m)}$ relative to the first density. Then this probability 
distribution will satisfy a LDP with the same rate $N$ and rate function obtained 
by adding $H(m)$ to the first rate function and subtracting a constant. 

Combining these two ingredients we obtain the third statement governing 
the large deviation properties of the non-equilibrium system started in the inverse 
temperature $\b$ and driven with inverse temperature $\b'$. 

\begin{thm}\label{thm:formJ2}
Denote by $P_{\b',\b, N}$ the law of the paths 
$(z_{N}(s))_{s\in [0,t]}$ of the magnetization for the Markov-chain with inverse
 temperature $\b'$ with initial condition distributed according 
to the Curie-Weiss measure $\mu_{\b',N}$. 

Then the measures $P_{\b,\b',N}$ satisfy 
%in $L_{\infty}[0,t]$ 
a large deviation principle 
with rate $N$  and rate function given by the Lagrange functional  
\begin{equation}\label{23}
\begin{split}
%\phi\mapsto-\frac{\b \phi(0)^2}{2}+I(\phi(0))+J_{\b'}(\phi)
\phi\mapsto H(\phi(0))+I(\phi(0))+J_{\b'}(\phi)
\end{split}
\end{equation}
\end{thm}

The knowledge of this compound rate function 
allows us to compute the large $N$ asymptotics 
of the probability to find the system in a final magnetization $m'$ at time $t$ 
by computing the value in the rate function in the minimizing path to $m'$. Note that  
this time the optimization is also over the initial point $m$. This minimizing 
path $\phi$ is found by solving the Euler-Lagrange equations with final condition $m'$ 
and an initial condition which is determined by another equation at the left-end point, which  
relates $\phi(0)$ and $\dot\phi(0)$ in an $\b$- and $\b'$-dependent way, as we will see.  
We also call the corresponding curve in  $(\phi(0),\dot\phi(0))$ {\em the curve 
of allowed initial configurations}.

\begin{cor}\label{thm:formJ3}
The conditional distribution of the initial magnetization $m$ 
taken according to the  law of the paths  $P_{\b',\b, N}$, 
conditioned to end in the final condition $m'$ at time $t$,  
satisfies a large deviation principle with 
rate $N$  and rate function given by   
\begin{equation}\label{39}
%E_{m'}(m,\b,\b')=-\frac{\b m^2}{2}+I(m)+
%\inf_{\genfrac{}{}{0pt}{3}{\phi:\phi(0)=m,}{\phi(t)=m'}} J_{\b'}(\phi)-\Const(m')
E_{m'}(m,\b,\b')=H(m)+I(m)+
\inf_{\genfrac{}{}{0pt}{3}{\phi:\phi(0)=m,}{\phi(t)=m'}} J_{\b'}(\phi)-\Const(m')
\end{equation}
\end{cor}

We are now ready to give our formula for the limiting conditional 
distributions of our model started at $\b$ and evolved with $\b'$.

\begin{thm} Fix $\b,\b',t,m'$. 
Suppose the constrained variational problem \eqref{23} for paths $\phi$ 
taken over the paths with fixed right endpoint $\phi(t)=m'$ 
has a unique minimizing path $s\mapsto m^*(s; m', t)$. 
 
Then the limiting probability kernels
of the time-evolved measure $\mu_{\b,\b',t;N}$
have a well-defined infinite-volume limit
$\g_{\b,\b',t}(\cdot|m')$ in the sense of (\ref{def:goodConf}) of the following form

\begin{equation}\label{25}
\begin{split}
\g_{\b,\b',t}(\eta_1|m')=\frac{\sum_{\s_1=\pm 1} e^{\s_1 \b m^*(0; m',t)}p_t( \s_1,\eta_1; m',t) }{
\sum_{\s_1,\tilde \eta_1=\pm 1}e^{\s_1 \b m^*(0; m',t)}p_t( \s_1,\tilde\eta_1; m',t) }
\end{split}
\end{equation}

Here $p_s( \sigma_1,\eta_1; m',t) $ is the probability to go from $\sigma_1\in \{-1,1\}$ at time $s=0$ 
to $\eta_1\in \{-1,1\}$ at time $s\leq t$ according 
to the Markov jump process on $\{-1,1\}$ which is defined by 
the time-dependent generator 
\begin{equation}\label{gabi}
\begin{split}
&L(s; m',t)f(\s_1)=c(\sigma_1,m(s; m',t))( f(-\sigma_1)-f(\sigma_1))\cr
\end{split}
\end{equation}
with rates which are obtained by substitution of the optimal path for 
the constrained problem for the empirical magnetization into the single-site 
flip rates. 
\end{thm}

{\bf Proof.} 

Take a sequence $\a_N \in \{-1,-1 + \frac{2}{N-1},\dots, 1-\frac{2}{N-1},1\}$ with the property $\lim\a_N=\a$ as $N\uparrow\infty$. 
We denote by $m_{N-1}(s)=\frac{1}{N-1}\sum_{i=2}^M \s(s)$ 
the empirical magnetization of the spins of site $2$ to $N$. To prove 
that the promised form for the limiting conditional probabilities is correct 
we must show that  
\begin{equation}\label{cp18}
\begin{split}
\lim_{N\uparrow \infty}\frac{\P_{\b,\b',N}(\s_1(t)=\eta_1| m_{N-1}(t)=\a_{N} )}{
\P_{\b,\b',N}(\s_1(t)=\eta'_1| m_{N-1}(t)=\a_{N} )}=\frac{\g_{\b,\b',t}(\eta_1|\a)}{\g_{\b,\b',t}(\eta'_1|\a)}
\end{split}
\end{equation}
where the r.h.s. is given by \eqref{25}. 

Let us abbreviate the whole path $(m_{N-1}(s))_{0\leq s \leq t}$ by the symbol $x$. 
Then, at finite $N$, a double conditioning gives us the identity of the form 
\begin{equation}\label{eq:condProbFrac}
\begin{split}
&\frac{\P_{_{\b,\b',N}}(\s_1(t)=\eta_1| m_{_{N-1}}(t)=\a_{N} )}{
\P_{_{\b,\b',N}}(\s_1(t)=\eta'_1| m_{_{N-1}}(t)=\a_{N} )}=\cr
&
\frac{\int \P_{_{\b,\b',N}}(d x | m_{_{N-1}}(t)=\a_{N} )\displaystyle{\sum_{\tilde \s_1=\pm 1}}
\P_{_{\b,\b',N}}(\s_1(0)=\tilde \s_1 | x) \P_{_{\b,\b',N}}(  \s_1(t)=\eta_1|   \s_1(0)=\tilde \s_1, x)
}{ \text{ the same with }\eta'_1 \text{ replacing }\eta_1
}
\end{split}
\end{equation}
We note that under our assumption on the solution of the constrained path large 
deviation principle  the distribution $\P_{\b,\b',N}(d x | m_{N-1}(t)=\a_{N})$  
concentrates exponentially fast on the trajectory $x^*:s\mapsto m^*(s; m', t)$ 
as $N$ tends to infinity. This collapses the outer expected value and simplifies 
the formula a lot. 
Next we have that whenever $x_N\rightarrow x^*$ we get 
$$\lim_{N\uparrow\infty}\P_{\b,\b',N}(\s_1(0)=\tilde \s_1 | x_N)=\frac{e^{\tilde \s_1 \b m^*(0; m',t)}}{
2\cosh(\b m^*(0; m',t))
}$$ 
Finally we have that the single-site Markov chain describing the time-evolution 
of the spin at site $1$, {\em conditional} on the 
path of the empirical mean of the other $N-1$ spins and its initial value at time $0$,  
converges to the Markov chain with deterministic but time-dependent generator \eqref{gabi}. 
The corresponding transition probabilities converge to 
the limiting expression from the theorem and we have 
\begin{equation}\label{eq:condProb}
\begin{split}
\lim_{N\uparrow\infty}
\P_{\b,\b',N}(  \s_1(t)=\eta_1|   \s_1(0)=\tilde \s_1, x_N)= p_t( \tilde \sigma_1,\eta_1; m',t) 
\end{split}
\end{equation}
This finishes the proof of \eqref{cp18}.$\Cox$

\bigskip 

%VARIATIONAL PROBLEM
\section{Phase-space geometry and multiple histories}
\subsection{Euler-Lagrange equations and curve of allowed initial configurations}
Fix $\b,\b',t,m'$. We look at the constrained variational problem \eqref{23} 
taken over the paths $\phi$  with $\phi(t)=m'$ with the aim to find  (the) minimizing path(s) $s\mapsto m^*(s; m', t)$. 
It is known in the calculus of variations 
%\cite{Elsgolts}
\cite{Gelfand-Fomin} (ch. 3, sect. 14)
that a necessary condition 
for an extremum is given by the corresponding Euler-Lagrange equation and an additional  
free left-end condition of the form 
\begin{equation}\label{eq:ELf}%Euler-Lagrange Formal
		\begin{array}{lcl}
		\frac{d}{ds}j_{\dot \phi}(\phi(s),\dot \phi(s)) -j_{\phi}(\phi(s),\dot \phi(s)) & = & 0 \text{   for all } s \in [0,t]\\
		j_{\dot \phi}(\phi(s),\dot \phi(s))- H_{\phi}(\phi(s)) - I_{\phi}(\phi(s)) \vert_{s=0} & = & 0 \\
		\phi(t) & = &m'
		\end{array}
\end{equation}
where $H$ denotes the initial Hamiltonian. Here we have dropped the subscript $\b'$ for 
the function $j(\phi(s),\dot \phi(s))$ and written subscripts to denote partial derivatives.  
It is straightforward to derive 
this set of equations by linear perturbation around the presumed  minimizing  
function $\phi(s)$ using a partial integration 
in the $s$ integral. For more details, see the Appendix. 

The first equation is a second order ODE which has two unknown parameters 
which have to be determined by the second and third equation. 
We call the curve described by the second equation which gives 
a condition between initial point and initial slope of the solution curve 
the {\em curve of the ``allowed'' initial configurations (\textsl{ACC})}.
We note that it is independent from the final value $m'$.  

Substituting the form of  $j(\phi(s),\dot \phi(s))$, we get after a small computation the equations 
\begin{equation}\label{eq:EL}%Euler-Lagrage 
		\begin{array}{lcl}
		\ddot m & = & 16 e^{2\b'm}\frac{\left((1+m)-e^{2\b'm}(1-m)\right)\left(1+\left(m^2-1\right)\b'\right)}{\left((1+m)+e^{2\b'm}(1-m)\right)^3}
		\\
		\dot m\Bigr|_{s=0} &= &g(m)\Bigr|_{s=0} 
		\\
		m(t)& = & m'
		\end{array}
\end{equation}
with the function 
\begin{equation}\label{eq:ACC}
g(m) =2e^{2\b'm}\frac{(1+m)-e^{2m(\b-\b')}(1-m)}{(1+m)+e^{2m\b'}(1-m)} 
\end{equation}
describing the curve of allowed initial configurations. 

Here we have written 
$m$ instead of $\phi(s)$, and the dot denotes time derivative w.r.t. $s$. 

\subsection{Typical paths for independent time-evolution}
Let us start with a discussion of the independent time-evolution. 
\\* {\bf (i)} For $\b'=0, \b=0$, the system becomes 
\begin{equation}\label{eq:c00}%Case both 0-s
		\begin{array}{rcl}
		\ddot m(s) & = & 4 m(s)						\\
		\dot m(s)\Bigr|_{s=0}  & = &2m(s)\Bigr|_{s=0}	\\
		m(t) & = &m'											\\
		\end{array}
		\end{equation}
and the solution becomes $m(s)=m'e^{2(s-t)}$. This describes how a curve which is conditioned to end 
in $m'$ away from zero is built up from the initial condition $m'e^{-2 t}$ close to zero. 
\\* {\bf (ii)} For independent dynamics $\b'=0$ and initial inverse temperature $\b \neq 0$ 
the  simplified system is
		\begin{equation}\label{eq:c0n}%Case 0 Non-0
		\begin{array}{rcl}
		\ddot m(s) & = & 4 m(s) \\
		\dot m(s)\Bigr|_{s=0} & = & e^{-2\beta m(s)}(1+m(s))-e^{2\beta m(s)}(1-m(s))\Bigr|_{s=0} \\
		m(t) & = & m'
		\end{array}
		\end{equation}
In this case the general solution is a linear combination of the $e^{\pm 2s}$. Looking 
at the right-end condition one gets 
$$m(s)=(m'-C_2 e^{2t})e^{2(t-s)}+C_2 e^{2s},$$
where $C_2$ is a constant and must be determined by the left-end condition. This can be done numerically.

It is possible to match the current approach with the one of \cite{KuLeNy} by plugging 
the solution curves with an initial condition 
$m(0)=m^*$ which are given by 
\begin{equation}\label{eq:consts}
	\begin{split}
			&m(s)=\frac{m^* e^{2t}-m'}{e^{2t}-e^{-2t}}e^{-2s}+\frac{m'-m^* e^{-2t}}{e^{2t}-e^{-2t}}e^{2s}, s \in \left[0,t\right]\cr
	\end{split}
\end{equation}
into the rate function and carrying out the time integral explicitly.  This gives 
\begin{equation}\label{eq:costIntegrated}
\begin{split}
&E_{m'}(m,\b,0)=H(m)+I(m)\cr
&+\frac{1}{4}\Biggl(4t+\ln\left[\frac{1-{m'}^2}{1-m^2}\right]
+2\left(m'\ln\left[\frac{R-C_1 e^{-2 t}+C_2 e^{2 t}}{1-m'}\right]-m \ln\left[\frac{R-C_1+C_2}{1-m}\right]\right)\cr
&+\ln\left[\frac{1-R-2 C_1 m' e^{-2 t}}{1+R-2 C_1 m' e^{-2 t}}\cdot\frac{1+R-2 C_1 m}{1-R-2 C_1 m}\right]\Biggr),\cr
&\text{where } R=\sqrt{1-4 C_1 C_2},\,\,
C_1=\frac{m e^{2t}-m'}{e^{2t}-e^{-2t}},\,\, C_2=\frac{m'-m e^{-2t}}{e^{2t}-e^{-2t}}\cr
\end{split}
\end{equation}
\begin{figure}[hb]
\centering
  \includegraphics[height=5.5cm]{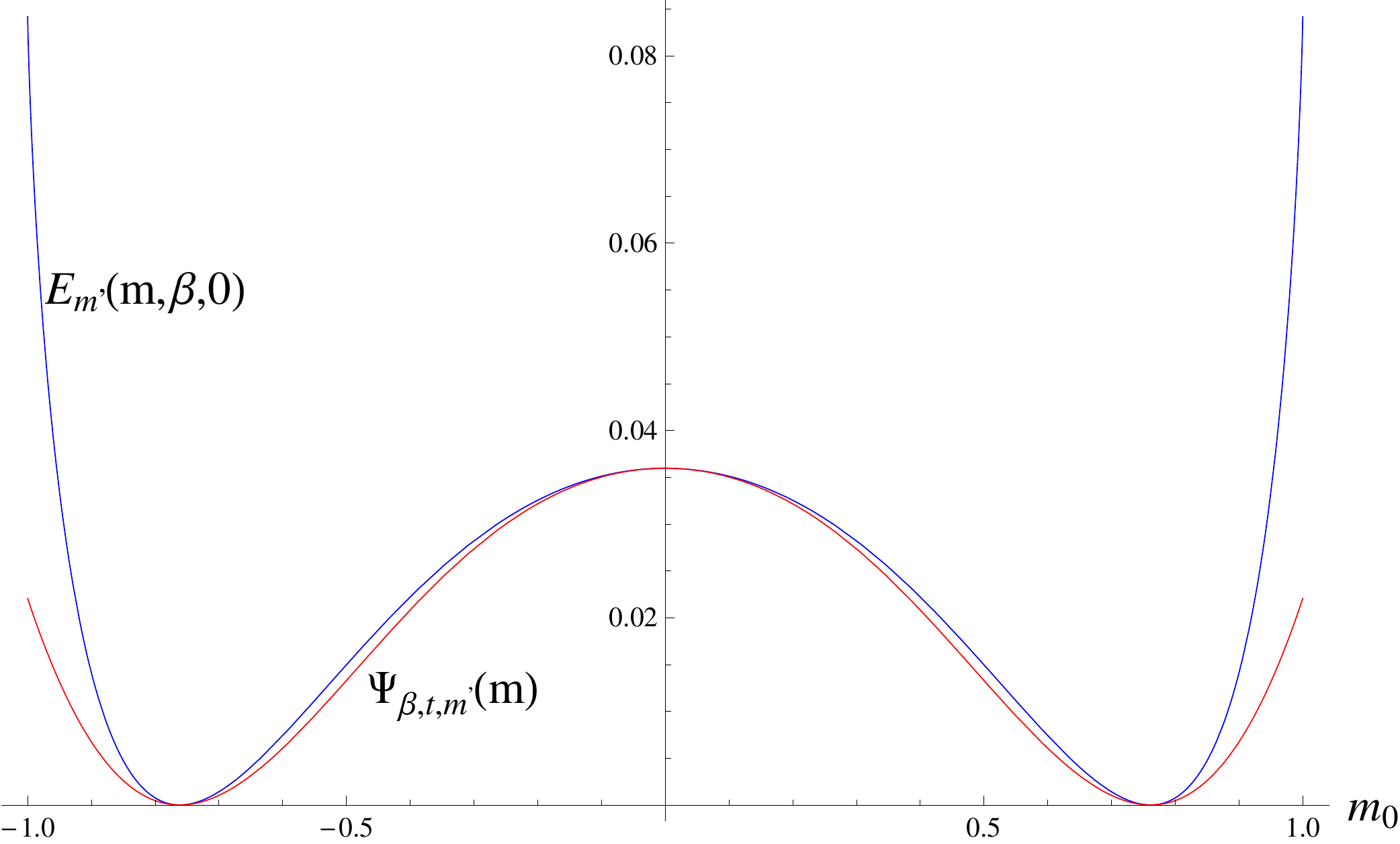}\\
	\caption{Cost functional $E_{m'}(m,\b,0)$ and known function $\Psi_{\b,t,m'}(m)$ for $\b'=0$, and $\b^{-1}\approx 1.744, t\approx 0.251$}
	\label{fig:Figger2}
\end{figure}

In the approach of \cite{KuLeNy} a related function called 
$\Psi_{\b,t,m'}(m)\label{eq:psi}$ was obtained by Hubbard-Stratonovitch transformation, 
whose minimizers with a given conditioning $(t,m')$ correspond to the most probable initial conditions.
This provides an opportunity to check if the results of the 
present analysis done via path large deviations coincide with the approach 
employing the function $\Psi_{\b,t,m'}(m)$. 

It is known that the functions $\Psi_{\b,t,m'}(m)$ \eqref{eq:psi} and $E_{m'}(m,\b,0)$ have the same set of extrema (see \cite{Opoku} in a more general context).
In Figure \ref{fig:Figger2} is the plot of these functions (after normalization to have zero as a minimum)
for the same set of parameters $(\b,m',t)$ which shows that the minima appear in fact 
at the same value.
\\*{\bf (iii)} Let us next turn to the case of interacting dynamics $\b'\neq 0$. In this case 
trajectories can only be obtained numerically. Before we go on, let us 
discuss in more detail the geometrical properties of the vector field and the allowed-configurations curve. 

\subsection{Geometric interpretation of Euler-Lagrange vector-field and curve of allowed initial configurations}
\begin{figure}[h]
\centering
 	\includegraphics[height=7cm]{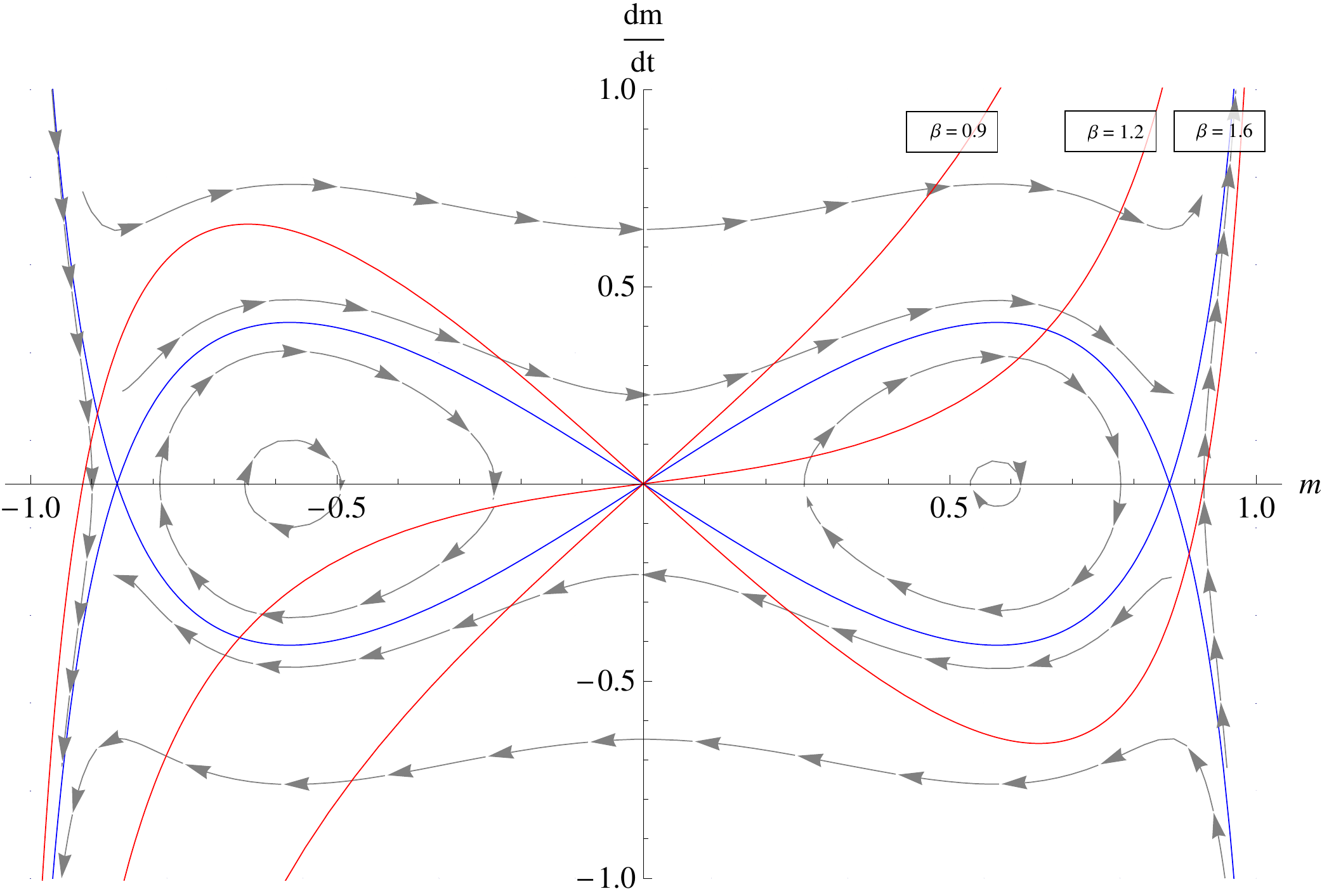}\\
	\caption{Phase portrait with level curves and \textsl{ACC}, $\b'=\frac32$}
	\label{fig:phasePortraitAllowed}
\end{figure}

Since the Euler-Lagrange density $j(\phi(s),\dot\phi(s))$ \eqref{eq:fullJ} does not
contain an explicit dependence on the time $s$, the generalized energy given by the Legendre transfrom of \eqref{eq:fullJ} is the system's first integral of motion 
\begin{equation}\label{eq:motionIntegral}
\begin{split}
&j(\phi(s),\dot \phi(s))-\dot \phi(s)j_{\dot \phi}(\phi(s),\dot \phi(s))=C\cr
\end{split}
\end{equation}
This can be rewritten as 
\begin{equation}\label{eq:FI1}
\begin{split}
&\frac{e^{4 \beta' m} (1-m)^2 \dot{m}^2+(1+m)^2 \dot{m}^2+2 e^{2 \beta ' m} \left(1-m^2\right) \left(8+\dot{m}^2\right)}{\left(1+e^{2 \beta' m} (1-m)+m\right)^2} = C\cr
\end{split}
\end{equation}
and explicitly solved for the velocity
\begin{equation}\label{eq:FI2}
\begin{split}
\dot m = \pm\sqrt{C+\frac{16 e^{2 \beta' m} (m^2-1)}{\left(1-e^{2\beta' m} (m-1)+m\right)^2}}
\end{split}
\end{equation}
Looking at the integral curves in phase space we get some geometric 
intuition.Let us go back to the notion of the \textsl{ACC} \eqref{eq:ACC} 
on which all possible ``allowed'' starting conditions lie.
In figure \ref{fig:phasePortraitAllowed} there are several \textsl{ACC}s drawn
which correspond to different values of $\b$, but the same value of the dynamical inverse 
temperature $\b'=\frac32$, which is reltively low. The production of discontinuities of the limiting 
conditional probabilities will be related to the time-evolution of the curve 
of allowed initial configurations under the Euler-Lagrange vector field, as we will describe now. 

Let us first give a definition of a bad quadruple of initial temperature, 
dynamical temperature, time, and final magnetization  in terms of dynamical-systems quantities.
We start by defining candidate quadruples making use of the Euler-Lagrange flow in the following way. 
\begin{defn}
The quadruple $(\b,\b',t,m_{\text{pb}})$ is called {\em pre-bad} iff there exists a pair 
$m_{0,1}\neq m_{0,2}$ of initial magnetizations s.t. the solution of the initial value 
problem of the Euler-Lagrange equations started in the corresponding points $(m_{0,1},g(m_{0,1}))$ and 
$(m_{0,2},g(m_{0,2}))$ on the allowed-configurations curve for $\b,\b'$ has the 
same magnetization value $m_{\text{pb}}$ at time $t$, that is 
$$m(t;m_{0,1},g(m_{0,1}))= m(t;m_{0,2},g(m_{0,2}))=m_{\text{pb}}$$
\end{defn}
While this first definition refers only to the existence of 
overhangs of the time-evolved allowed-configurations curve,  
the next definition involves also the value of the cost (the large deviation functional 
together with the punishment term), which makes it much more restrictive. 
\begin{defn}
The pre-bad quadruple $(\b,\b',t,m_{\text{bad}})$ is called {\em bad} iff the two different
paths started at the corresponding $m_{0,1}\neq m_{0,2}$ are both minimizers for the cost, i.e.  
$$E_{m_{\text{bad}}}(m_{0,1},\b,\b')=E_{m_{\text{bad}}}(m_{0,2},\b,\b')
=\inf_{m} E_{m_{\text{bad}}}(m,\b,\b')$$
\end{defn}
We will exploit both definitions both to gain geometric insight as well as numerical results. 
The important connection to non-Gibbsian behavior of the time-evolved measure lies in the fact that 
$m_{\text{bad}}$ of a bad quadruple will (generically) be a bad configuration for $\g_{\b,\b',t}(\cdot |m)$.  
Indeed, to see this, let us go back to the explicit expression of the limiting conditional probabilities, given by
\begin{equation}\label{255}
\begin{split}
\g_{\b,\b',t}(\eta_1|m')=\frac{\sum_{\s_1=\pm 1} e^{\s_1 \b m^*(0; m',t)}p_t( \s_1,\eta_1; m',t) }{
\sum_{\s_1,\tilde \eta_1=\pm 1}e^{\s_1 \b m^*(0; m',t)}p_t( \s_1,\tilde\eta_1; m',t) }
\end{split}
\end{equation}
Note that the function $m^*(0; m',t)$ is not well defined for $m'=m_{\text{bad}}$ itself since at time $t$ 
there are two minimizing paths available, one from $m_{0,1}$ to $m_{\text{bad}}$ and 
one from $m_{0,2}$ to $m_{\text{bad}}$. Varying however {\em around} $m_{\text{bad}}$ the 
paths will become unique and we might select the minimizing paths (and hence their initial points) 
by approaching the bad configuration 
from the right or left, obtaining (say) 
$\lim_{m'\downarrow m_{\text{bad}}} m^*(0; m',t)=m_{0,1}$ and 
$\lim_{m'\uparrow m_{\text{bad}}} m^*(0; m',t)=m_{0,2}$. 
Note that we also expect that (generically) 
$\lim_{m'\downarrow m_{\text{bad}}}p_t( \s_1,\tilde\eta_1; m',t)\neq \lim_{m'\uparrow m_{\text{bad}}}p_t( \s_1,\tilde\eta_1; m',t)$. This follows since the $p_t$ are probabilities for two different single-particle Markov chains, 
one depending on the path starting from $(m_{0,1},g(m_{0,1}))$, the other one 
on the path  starting from $(m_{0,2},g(m_{0,2}))$. 
We note that, knowing the paths entering the $p_t$'s, an explicit formula 
for $p_t$ in terms of time-integrals can be written, and so, given (numerical) knowledge 
of the minimizing path, the $\g_{\b,\b',t}(\eta_1|m')$ can be obtained by simple integrations. 
Unless these two discontinuities compensate each other (which is generically not happening 
and which can be quickly checked by numerics) we will have that  
$\lim_{m'\downarrow m_{\text{bad}}}
\g_{\b,\b',t}(\eta_1|m')\neq \lim_{m'\uparrow m_{\text{bad}}} \g_{\b,\b',t}(\eta_1|m')$. 
Consequently the model will be non-Gibbs at the time $t$.

Conversely, if $(\b,\b',t,m_{\text{pb}})$ is not {\em bad}, then 
$m'\mapsto \g_{\b,\b',t}(\eta_1|m')$ is a continuity point. This follows since in that case 
all $m'$-dependent terms in \eqref{255} deform in a continuous way. So the absence of bad points 
(and a fortiori the absence of pre-bad 
points) implies Gibbsianness at $(\b,\b',t)$. 

\subsection{Time-evolved allowed initial configurations} 

We just saw that non-Gibbsianness is produced by multiple histories which means
in other words the production of overhangs in the time-evolved curve of allowed initial
configurations. To get an intuition for this let us discuss  the regions 2) and 3) of the main
Theorem in more detail.
Let us begin with the phase-space picture for the non-interacting dynamics $\b'=0$.
We are starting with the region 2a) of non-symmetry-breaking non-Gibbsianness
i.e. $\frac{2}{3}=\b^{-1}_{\text{SB}}(\b'=0)\leq\b^{-1}<\min\{{\b'}^{-1},1\}=1$.

\begin{figure}[htb]
\centering
\begin{tabular}{l l}
 \includegraphics[height=4.5cm]{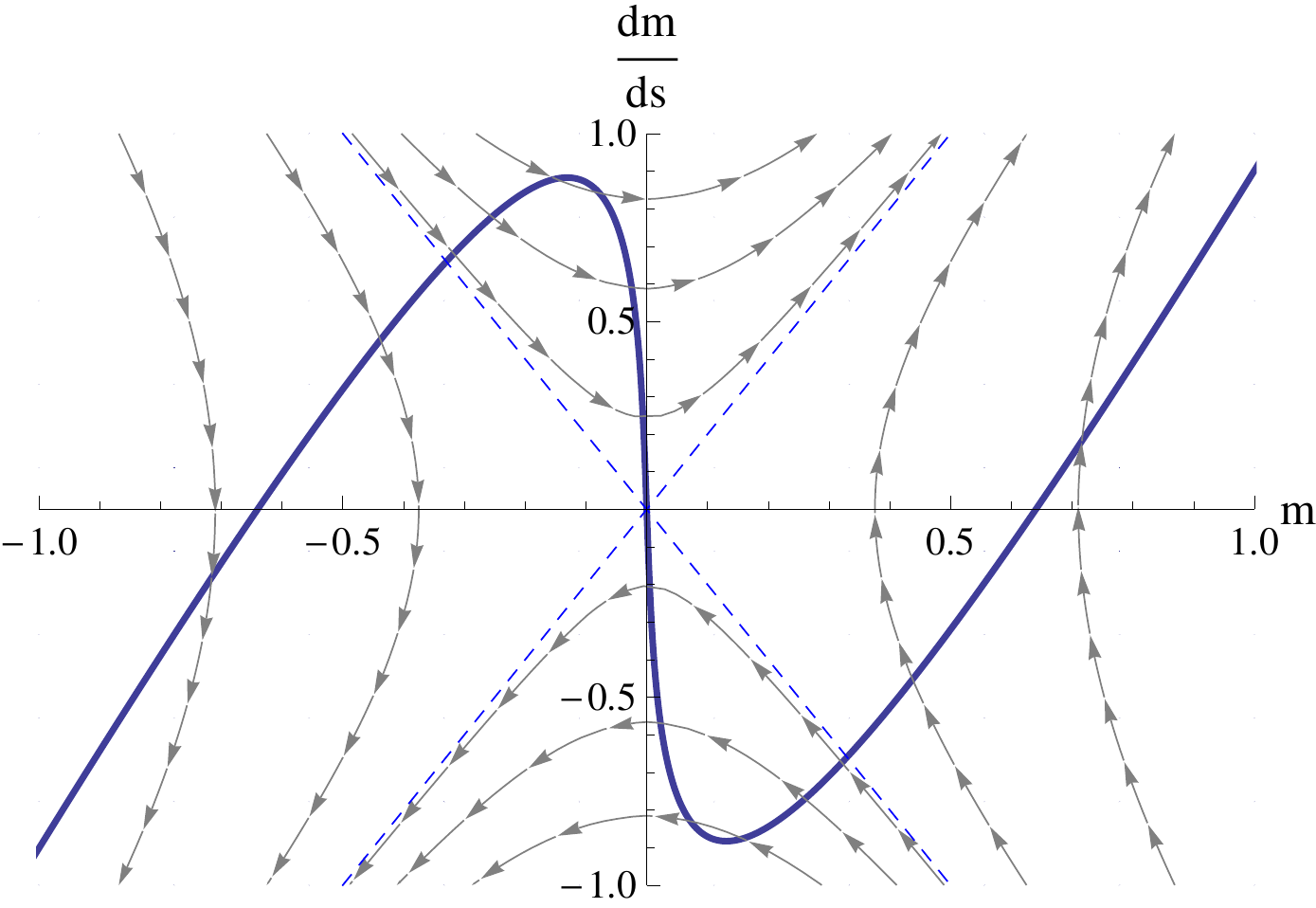}&
 \includegraphics[height=4.5cm]{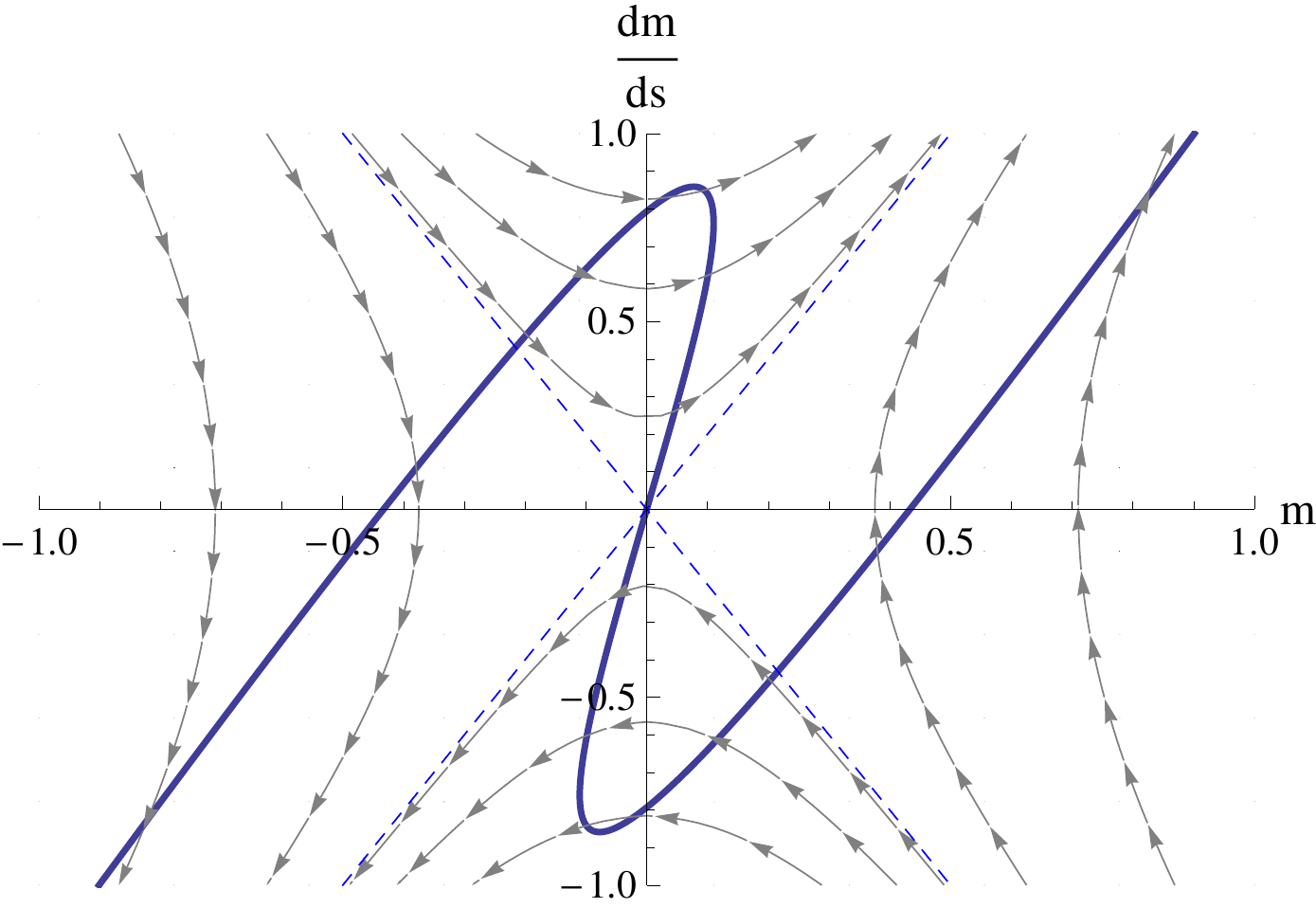}\\
\end{tabular}
\caption{Non-symmetry-breaking mechanism, $\b'=0, \b^{-1}=0.8$}
\label{fig:nsb-mech}
\end{figure}

The time-evolved allowed-configurations curve for 
$t = t_{\text{nGS}}(\b,\b'=0)$ is shown at the left plot of Figure \ref{fig:nsb-mech} where it acquires a vertical slope at zero. The right plot shows  the time-evolved allowed-configurations curve for 
$t > t_{\text{nGS}}(\b,\b'=0)$ where it has two symmetric overhangs. In particular $(\b,\b'=0,t,m'=0)$ is pre-bad. 
It is also bad, since the preimages of the upper and lower time-evolved allowed-configurations curve 
which intersect the vertical axis have paths with the same cost, by the symmetry of the model. 
Note that $(\b,\b'=0,t,m')$ is pre-bad for a whole interval of values of $m'$, but (as the study of the 
cost shows and as it was proved in \cite{KuLeNy}) there are no other bad points. 
We note that $m'=0$ is easily checked to be indeed a bad configuration (discontinuity point) of 
$\g_{\b,\b'=0,t}(\cdot |m')$ 
since there are no cancellations of discontinuities in this case, as we will explain now: Indeed, 
$p_t( \s_1,\tilde\eta_1; m',t)$ does not depend on the trajectory of the empirical magnetization 
and is given by the independent spin-flip at the site $1$ between plus and minus with rate $1$, 
\begin{equation}
\begin{split}
\g_{\b,\b'=0,t}(\eta_1|m')=\frac{\sum_{\s_1=\pm 1} e^{\s_1 \b m^*(0; m',t)}p_t( \s_1,\eta_1) }{
\sum_{\s_1,\tilde \eta_1=\pm 1}e^{\s_1 \b m^*(0; m',t)}p_t( \s_1,\tilde\eta_1) }
\end{split}
\end{equation}
where $p_t(+,+)=\frac{1}{2}(1+e^{-2 t})$, and 
$p_t(+,+)=p_t(-,-)=1-p_t(+,-)=1-p_t(-,+)$. So, a discontinuity under variation 
of $m'$ is entering the formula only through $m^*(0; m',t)$, and hence 
$m'\mapsto \g_{\b,\b'=0,t}(\eta_1|m')$ is discontinuous iff $m\mapsto m^*(0; m',t)$ is 
discontinuous. 

\begin{figure}[htb]
\centering
\begin{tabular}{l l}
 \includegraphics[height=4.5cm]{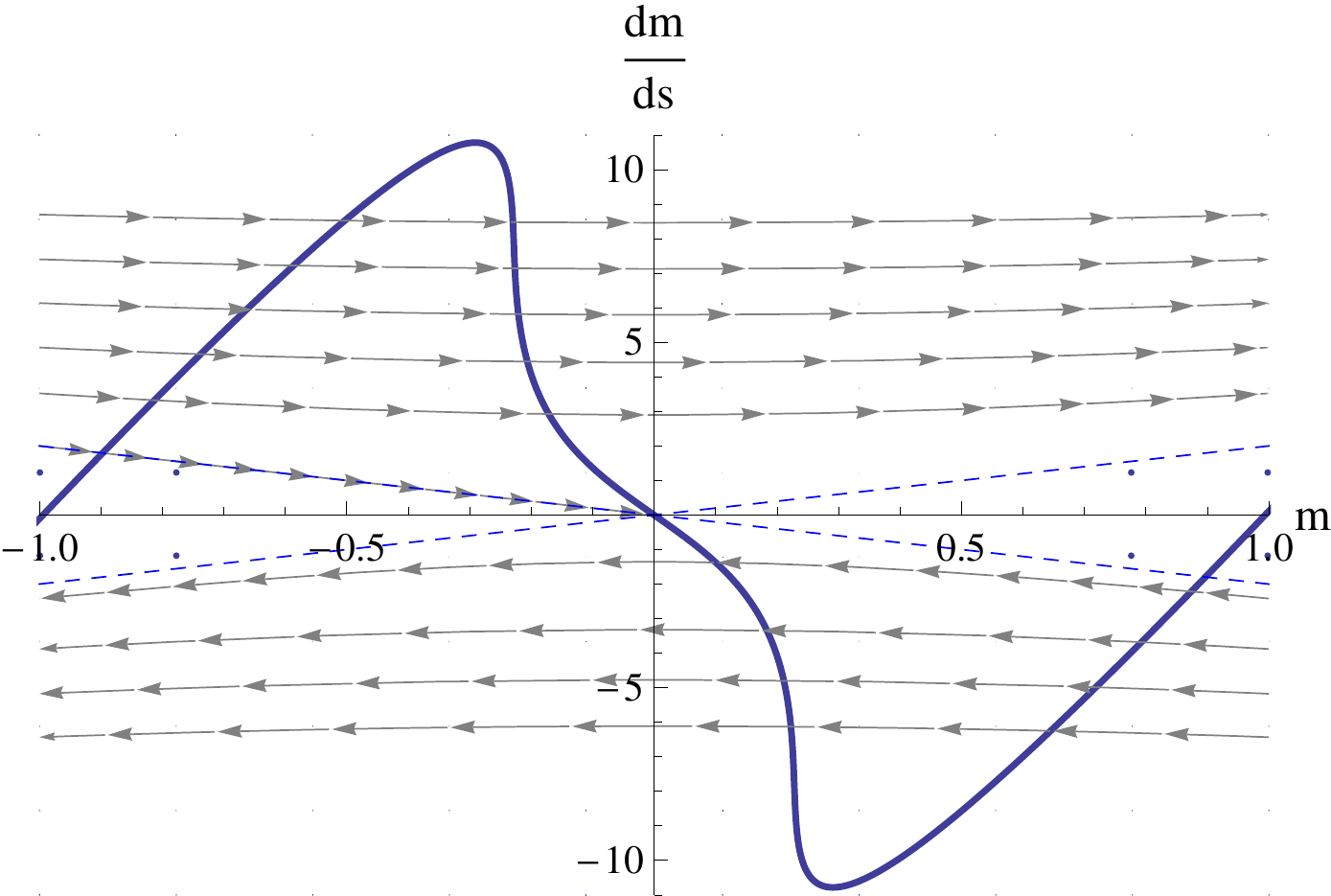}&
 \includegraphics[height=4.5cm]{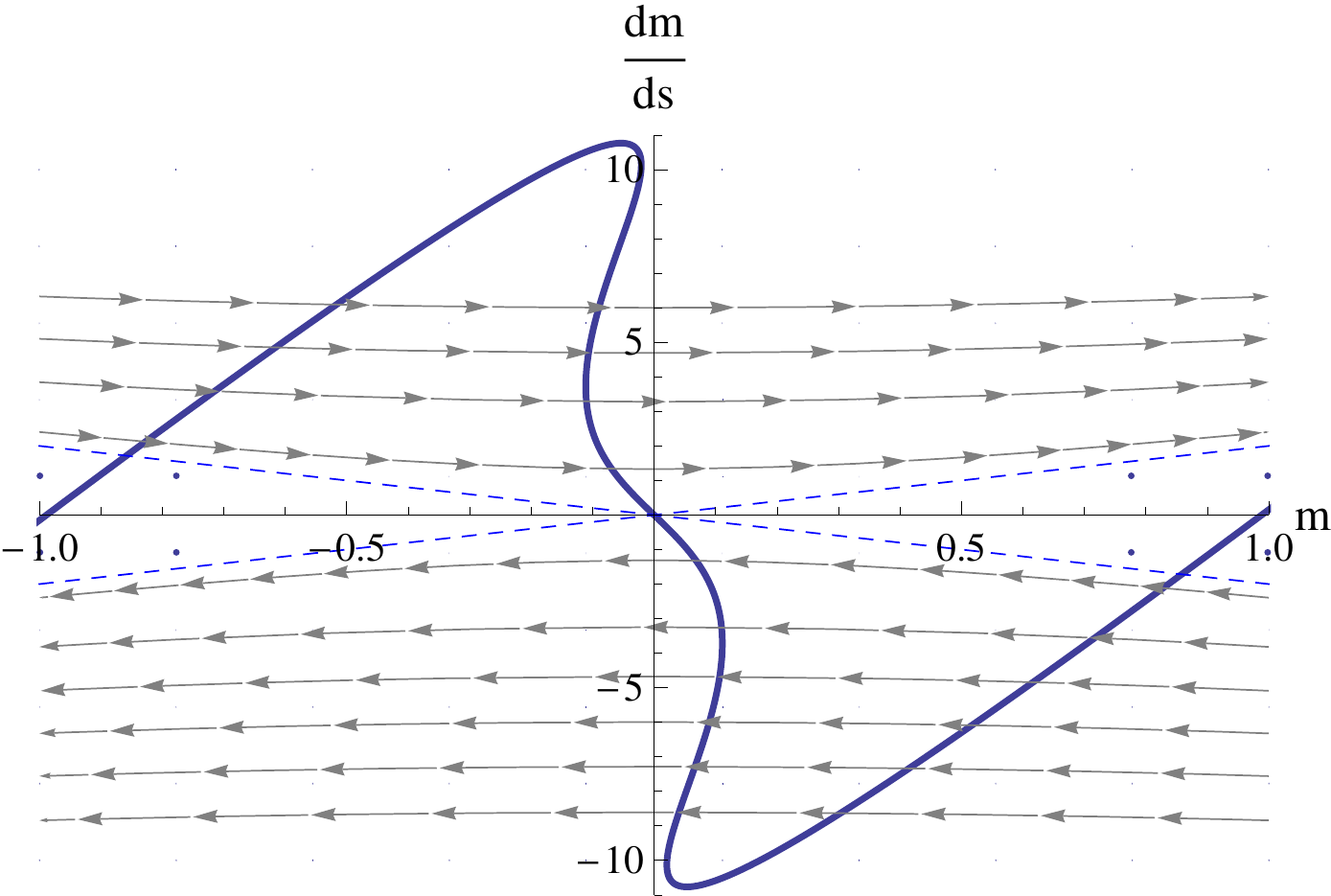}\\
\end{tabular}
\caption{Symmetry-breaking mechanism, $\b'=0, \b^{-1}=0.4$}
\label{fig:sb-mech}
\end{figure}
Let us now look at region 2b) of symmetry-breaking non-Gibbsianness i.e.  
$\b^{-1}<\b^{-1}_{\text{SB}}(\b'=0)$

The left plot of Figure \ref{fig:sb-mech} shows the time-evolved allowed-configurations curve 
at $t = t_0(\b,\b'=0)$ where it acquires a vertical slope away from zero.   
The right plot shows  the time-evolved allowed-configurations curve for 
$t_0(\b,\b')<t<t_1(\b,\b')$ where it has two symmetric overhangs away from zero. 
This means that $(\b,\b'=0,t,m')$ is pre-bad for a whole range of values of final magnetizations 
$m' $. Due to the lack of symmetry it is not clear to identify in the picture 
which of the $(\b,\b'=0,t,m')$'s will be bad. It turns out that it is precisely one 
such value  $(\b,\b'=0,t,m_c)$, and this can be found looking numerically at the cost. 

Perturbations of these pictures stay true for ${\b'}^{-1}> 1$, where they describe 
the only mechanism of non-Gibbsianness. 
Perturbations of these pictures {\em also} stay true for ${\b'}^{-1}<1$, but then there is also 
the Region 3 of the main theorem which describes the  cooling from an initial low temperature.
We choose $\frac23= {\b'}^{-1}<\b^{-1}= 0.85<1$. 
Then the vector field has periodic orbits which are intersected by the allowed-configurations curve, 
and the time-evolution will create overhangs and smear out the allowed-configurations curve over time. 

\begin{figure}[htb]
\centering
\begin{tabular}{l l}
 \includegraphics[height=4.5cm]{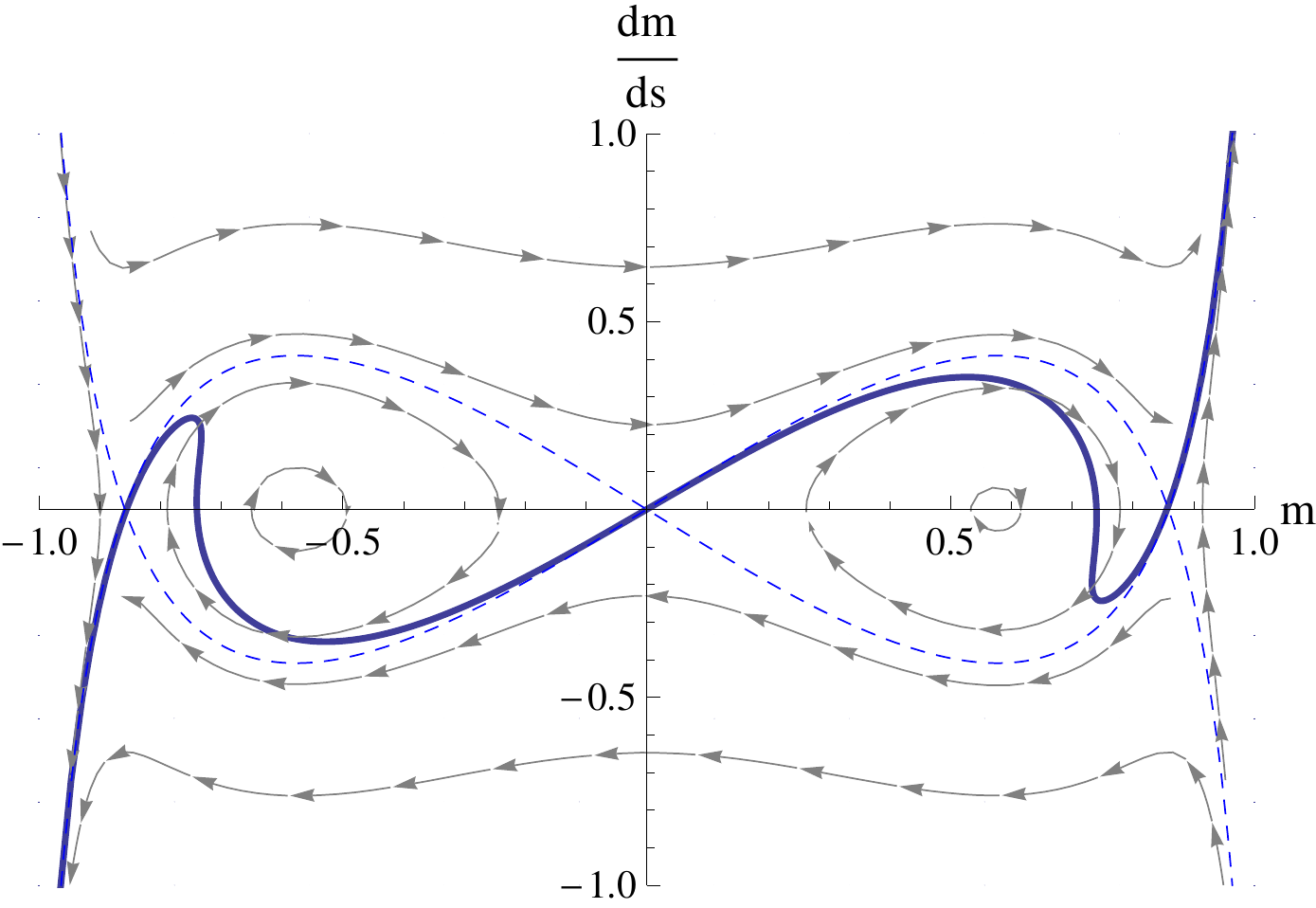}&
 \includegraphics[height=4.5cm]{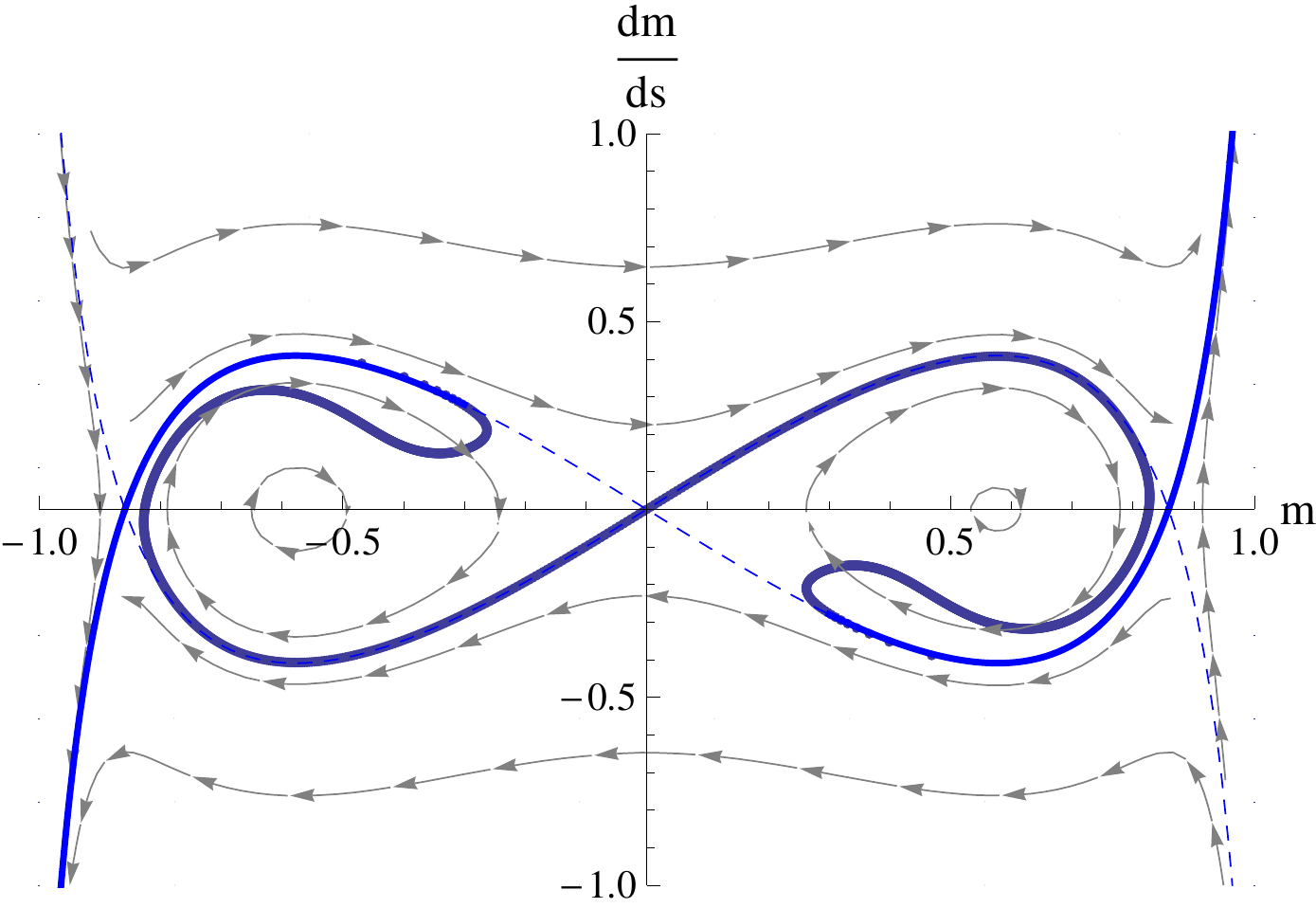}\\
\end{tabular}
\caption{Non-Gibbsianness by periodicity, ${\b'}^{-1}=\frac23, \b^{-1}=0.85$}
\label{fig:periodic-mech}
\end{figure}

The left plot of Figure \ref{fig:periodic-mech} shows the time-evolved allowed-configurations curve 
at $t = t_{\hbox{per}}(\b,\b')$ where it acquires a vertical slope away from zero inside the area of periodic motion. 

The right plot shows  the time-evolved allowed-configurations curve for a time 
$t>t_0(\b,\b')$ where it has overhangs. Again, from the interval of pre-bad points, the bad point 
has to be selected by looking at the cost. 
When time gets larger more overhangs are created and the trajectory is smeared out.  
The corresponding potential function $m\mapsto E_{m'}(m,\b,\b')$ will acquire more and more local extrema 
as $t$ increases. Then, by finetuning of the $m'$ while keeping the $\b,\b',t$ fixed, 
equality of the depths of the two lowest minima can be achieved. Since the number of available minima 
is increasing with $t$ we conjecture that there will be also an increasing number  
of bad $m'$s which becomes dense as $t$ increases. To prove this conjecture however, more investigation is needed. 

\subsection{Emergence of bad points as a function of time}
The notion of a bad point can be viewed from two different standpoints.
A pre-bad point in the time-space diagram is a point where two (or more) histories collide. 
If the costs computed along these paths are equal, then a pre-bad point is a bad point. In the phase space this  means that the phase flow transported two (or more) points originally lying on the curve of allowed initial configurations to the same space-position within equal time 
but with different speeds. Two (or more) points have the same space-position if their projections to the $m$-axis are equal, as seen in Figures \eqref{fig:nsb-mech}, \eqref{fig:sb-mech}, and \eqref{fig:periodic-mech}.
How can we identify analytically the first time $t$ where 
time-evolved initial points from the curve of allowed initial configurations 
will obtain the same projection to the $m$-axis? As intuition suggests one has to look when the
transported curve of allowed configurations aquires a vertical slope for the first time. 
This discussion brings us to the following computation.

%How can we identify analytically the first time $t$ where 
%time-evolved initial points from the curve of allowed initial configurations 
%will obtain the same projection to the $m$-axis? Such situations were depicted 
%in the left pictures of Figures \eqref{fig:nsb-mech}, \eqref{fig:sb-mech}, and \eqref{fig:periodic-mech}. 

Writing $v=\dot m$ for the velocity, let us consider the flow 
$m(t; m_0,v_0)$, $v(t; m_0,v_0)$
of our system under the Euler-Lagrange equations,
\begin{equation}\label{eq:genEL}
		\begin{array}{rcl}
		\dot m & = & v \\
		\dot v & = & f_{\b'}(m)
		\end{array}
\end{equation}
We take the curve of allowed initial configurations 
to be transported by the flow $v_0=g_{\b,\b'}(m_0)$ where 
we write in short $f=f_{\b'}$ and $g=g_{\b,\b'}$. 
We are then interested in the projections to the $m$-axis of 
the time-evolved curves in phase space, 
that is the curves $m_0 \mapsto m(t; m_0,g(m_0))$, as they evolve with $t$. 
Restricted to suitable neighboorhoods this curve becomes a function,  and we view 
it as a potential function with state variable $m_0$ and 
parameter $t$ (keeping also $\b,\b'$ as fixed parameters.) 

Doing so we see that the derivatives of the flow w.r.t. the initial conditions obey at the threshold time $t$ that 
\begin{equation} 
 \begin{split}\label{eq:emergence}
0&=	F_{\b',\b}(t,m_0):=\frac{d m(t; m_0,g(m_0))}{d m_0}
		=\frac{\partial m(t; m_0,v_0)}{\partial m_0}+ \frac{\partial m(t; m_0,v_0)}{\partial v_0}g'(m_0)\cr
0&=\frac{d^2 m(t; m_0,g(m_0))}{(d m_0)^2}		
\end{split}
\end{equation}
The first equation means that in the $(m,v)$ plane the time-evolved curve will obtain a vertical slope 
which is clear 
by the interpretation of the variable 
$m_0$ as a parametrization of the curve of allowed initial configurations. 

Moreover we have that the second derivative will also vanish, since a minimum and a maximum 
of $m_0 \mapsto m(s; m_0,g(m_0))$ collide for $s\downarrow t$, in a {\em fold bifurcation}. 

\subsection{The threshold time for non-symmetry-breaking non-Gibbsianness for dependent dynamics}
We can use these equations to obtain quantitative information about the threshold time 
for non-symmetry-breaking non-Gibbsianness also for dependent dynamics. For this it suffices 
to look at the dynamics locally around the origin $(m,\dot m)=(0, 0)$ in phase space which is a stationary point for the 
dynamics independently of $\b'$. 

Linearizing $f_{\b'}$%, or, which is the same, taking quadratic terms, 
we get
\begin{equation} 
\left(
\begin{array}{c}  
      \dot m \\
      \dot v 
\end{array}
\right)
=
\left(
\begin{array}{cc}  
      0 & 1 \\
      4(1-\b')^2 & 0 
\end{array}
\right)
\left(
\begin{array}{c}  
      m \\
      v 
\end{array}
\right)
\end{equation}
Corrections are only of third order. 
The eigenvalues of the matrix are $\lambda_{1,2}=\pm 2(1-\b')$, these eigenvalues are real and
have different signs, so $(m,\dot m)=(0,f_{\b'}(0))=(0,0)$ is a saddle point.
This ensures that the nature of solutions close to $(0;0)$ stays the same whatever
$\b'$ is taken.

Let us now discuss the phase flow around the origin $(0,0)$. 
At this point non-Gibbsianness without symmetry-breaking occurs, by the following argument. 
Suppose a symmetric pair of initial conditions 
$(m_0,v(m_0))$ and $(-m_0,v(-m_0))=(-m_0,-v(m_0))$ 
is given which has the same time-evolved magnetization $0$ at time $t$. This corresponds to the fact that the transported 
curve will have overhangs at the points $(0,v_{1}(m))$ and $(0,-v_{1}(m))$. If we look at the phase 
portraits of the dynamics as a function of time we see that for times  larger than but very close to the 
first time where this occurres the speed $v_{1}(m)$ will be very close to $0$. It converges to 
$0$ when $t$ approaches the transition time for Gibbsianness. Indeed 
the whole path was evolving in an arbitrarily small neighborhood of the origin and 
hence it suffices to look at the linearized dynamics. We also note that there is no need
to look at the cost functional in this case, due to the symmetry of the paths. 
As time becomes larger than the transition-time (as in the right 
picture of Figure 4) the intersection 
points of the time-evolved curve with the vertical axis will move away from zero and 
so it would not be sufficient to use the linearization of the dynamics to compute the relation 
between bad magnetization values and time. 

Clearly the general solution of the linearized system is 
\begin{equation} 
\begin{split}
	x(s)=C_1 e^{-2(1-\b')s}+C_2 e^{2(1-\b')s}
\end{split}
\end{equation}
%This knowledge allows us to write the phase flow explicitly by the putting
Putting the initial condition to be $(m_0,v_0)$ the phase flow becomes 
\begin{equation} 
 \begin{split}
	&m(s; m_0,v_0)= \frac{2 (1-\b')m_0-v_0}{4(1-\b')}e^{-2(1-\b')s} + \frac{2(1-\b')m_0+v_0}{4(1-\b')}e^{2(1-\b')s}     \cr
	&v(s; m_0,v_0)= \frac{v_0-2(1-\b')m_0}{2}e^{-2(1-\b')s} + \frac{v_0 + 2(1-\b')m_0}{2}e^{2(1-\b')s} \cr
\end{split}
\end{equation}
Computing the function $F_{\b',\b}(t,m_0)$ \eqref{eq:emergence} for this phase flow and setting it to zero, we solve it w.r.t. time $t$ and get
\begin{equation}\label{eq:timeNG}
\begin{split}
	t%=\frac{1}{4(1-\b')}\ln\frac{g'(m_0)-\sqrt{f'(0)}}{g'(m_0)+\sqrt{f'(0)}}
	 = \frac{1}{4(1-\b')}\ln\frac{g'(m_0)-2(1-\b')}{g'(m_0)+2(1-\b')}
\end{split}
\end{equation}

Putting $m_0=0$ we obtain from this for the transition time
\begin{equation}\label{eq:timeNGspec}
\begin{split}
	t=\frac{1}{4(1-\b')}\ln\frac{\b'-\b}{1-\b}
\end{split}
\end{equation}
By setting $\b'=0$ for the independent evolution 
in the last expression, the result  $t=\frac14 \ln(1-\b^{-1})$  given in \cite{KuLeNy} is reproduced.
We note that the transition time given by formula \eqref{eq:timeNGspec} 
is positive only in the case when $\b>1$. 
\begin{figure}[t]
\centering
\includegraphics[height=7.5cm]{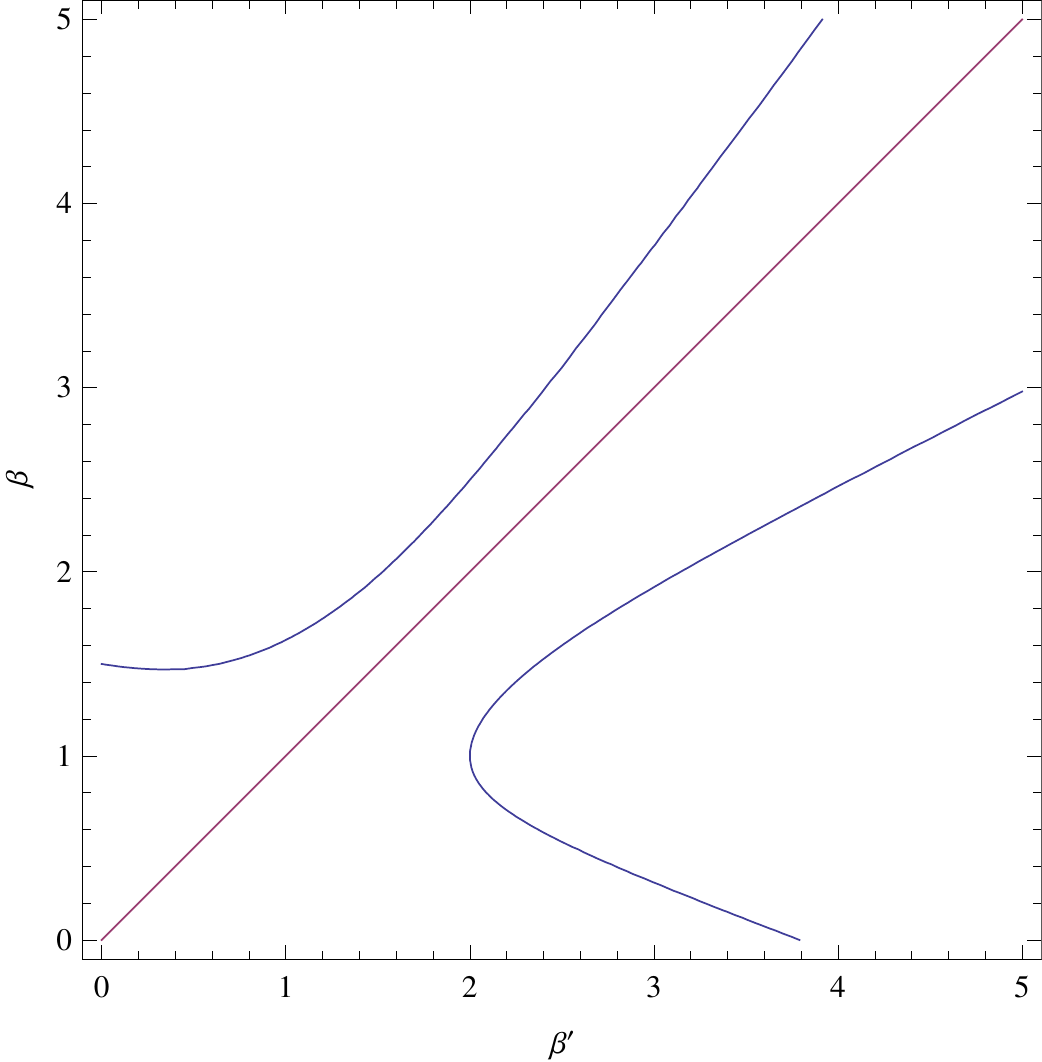}
\caption{The upper branch shows the symmetry-breaking 
inverse temperature $\b_{\text{SB}}$ as a function of $\b'$}
\label{fig:bonbp}% beta on beta prime
\end{figure}

To identify for which temperature-values the phenomenon 
of non-Gibbsianness without symmetry-breaking ends, 
let us look when the function \eqref{eq:timeNGspec} starts having several minima. In order to do this we compute the second derivative of \eqref{eq:timeNGspec} and put it equal to zero. 
This results in the equation
\begin{equation}\label{eq:bbp}%beta-beta-prime
 \begin{split}
	4 \b^3 + 12 \b \b' - 6 \b^2 (1 + \b') - \b' (3 + 3\b' - {\b'}^2)=0
\end{split}
\end{equation}
In the independent-dynamics case $\b'=0$ 
we get exactly $\b=\frac32$, which was already found in the paper \cite{KuLeNy}.

\subsection{Cooling and non-Gibbsianness by periodic orbits}

Let us specialize to the case of a low-temperature dynamics $\b' >1$.  
In that case the phase space decomposes into the 
areas of periodic and non-periodic dynamics. The separatrix is given by \eqref{eq:FI2} with $C=4$.
\begin{equation}\label{eq:sepx}
\begin{split}
f_{\pm}(m)=\pm2\frac{(1+m)-e^{2\b'm}(1-m)}{(1+m)+e^{2\b'm}(1-m)}
\end{split}
\end{equation}
Note that the curve $f_{+}(m)$ coincides with the curve of the ``allowed'' configurations \eqref{eq:ACC}  when $\b'=\b$. This means that it will be stable under the phase flow in that case. 
In particular the time-evolved 
curve will not acquire overhangs which corresponds to the fact that the time-evolved 
measure will be invariant under the dynamics and the model Gibbs. 

Note also that the negative branch of the 
separatrix % (``negative'' branch---$f_-$) 
coincides with the right-hand side of the ODE describing the unconstrained typical evolution 
\eqref{eq:deterministicR} and so the intersection point with the $m$-axis is given 
given by the
biggest solution of the ordinary mean-field equation $m=\tanh(\b' m)$. Let us first 
concentrate of the existence of pre-bad points, that is different initial points of the allowed-configurations curve leading to the same projection to the $m$-axis after time $t$. 

Now multiple overhangs are created if the allowed curve of initial configurations 
intersects the periodic motion area, as seen in Figure 6. 
Indeed this part of the curve will perform periodic 
motion and while doing so it will acquire more and more overhangs, filling out the part 
of the periodic motion area which is bounded by its extremal value of the integral of motion over 
time.  It is now interesting to note for which temperatures this phenomenon can happen 
and this is the content of the following theorem. 
 
\begin{thm}[Non-Gibbsianness by periodicity] \label{thm:NGperiodic}
Suppose $\b'>1$ and let  $m_1^* (m_2^*)$ be the biggest solution of the mean-field equation
for $\b' (\b)$.
Then the following is true. 
\begin{enumerate}
	\item 
	if $1<\b<\b'$(or equivalently $0<m_2^*<m_1^*$) holds then	
	\begin{itemize}
		\item The curve of allowed initial configurations for $\b,\b'$ has non-zero intersection with the 
		(open) periodic motion area in phase phase for $\b'$. 
		%		A periodic motion area is identified by interval $(-m_1^*,0)\cup(0,m_1^*)$
		\item Consequently there exists a threshold time $t_{\text{per}}(\b,\b')$ such that 
		for all $t>t_{\text{per}}(\b,\b')$ there exists pre-bad $(\b,\b',t,m')$s. 	
	\end{itemize}
	\item	if $1<\b<\b'$ fails, there is either no periodic motion areas, or 
				the curve of allowed-configurations has no intersection with them. 
\end{enumerate}
\end{thm}

{\bf Proof. }
Denote  $f=f_-$ (here we take the branch which bounds the periodic motion area from above), and the curve of the ``allowed'' configurations by $g(x)$ (here $x$ is used instead of $m$) so that we have  
\begin{equation}
\begin{split}
&f(x)=-2\frac{(1+x) - e^{2 x \b'}(1-x)}{(1+x) + e^{2 x \b'}(1-x)},\cr
&g(x)=2e^{2\b'x}\frac{(1+x)-e^{2x(\b-\b')}(1-x)}{(1+x)+e^{2x\b'}(1-x)}
\end{split}
\end{equation}
Previously it was mentioned that periodic motion arises only in the case $\b'>1$, and 
so we will consider this along the proof, also w.l.g. we say that $x>0$.
Let us show what the condition $1<\b<\b'$ means and its equivalence to $0<m_2^*<m_1^*$.
First, we put $f(x)=0$ to determine the right border of the periodic motion area, and we get
that it's given by the equation $$(1+x)-e^{2\b'x}(1-x)=0,$$ which is equivalent 
to the mean-field equation for $\b'$.
Second, consider $f(x)=g(x)$ to determine their intersection point.
This is simply $$(1+x)-e^{2\b x}(1-x)=0,$$ which is again the same mean-field equation, but for
$\b$, where $m^*_2$ has the same meaning as before.

\begin{figure}[htb]
\centering
\includegraphics[height=6cm]{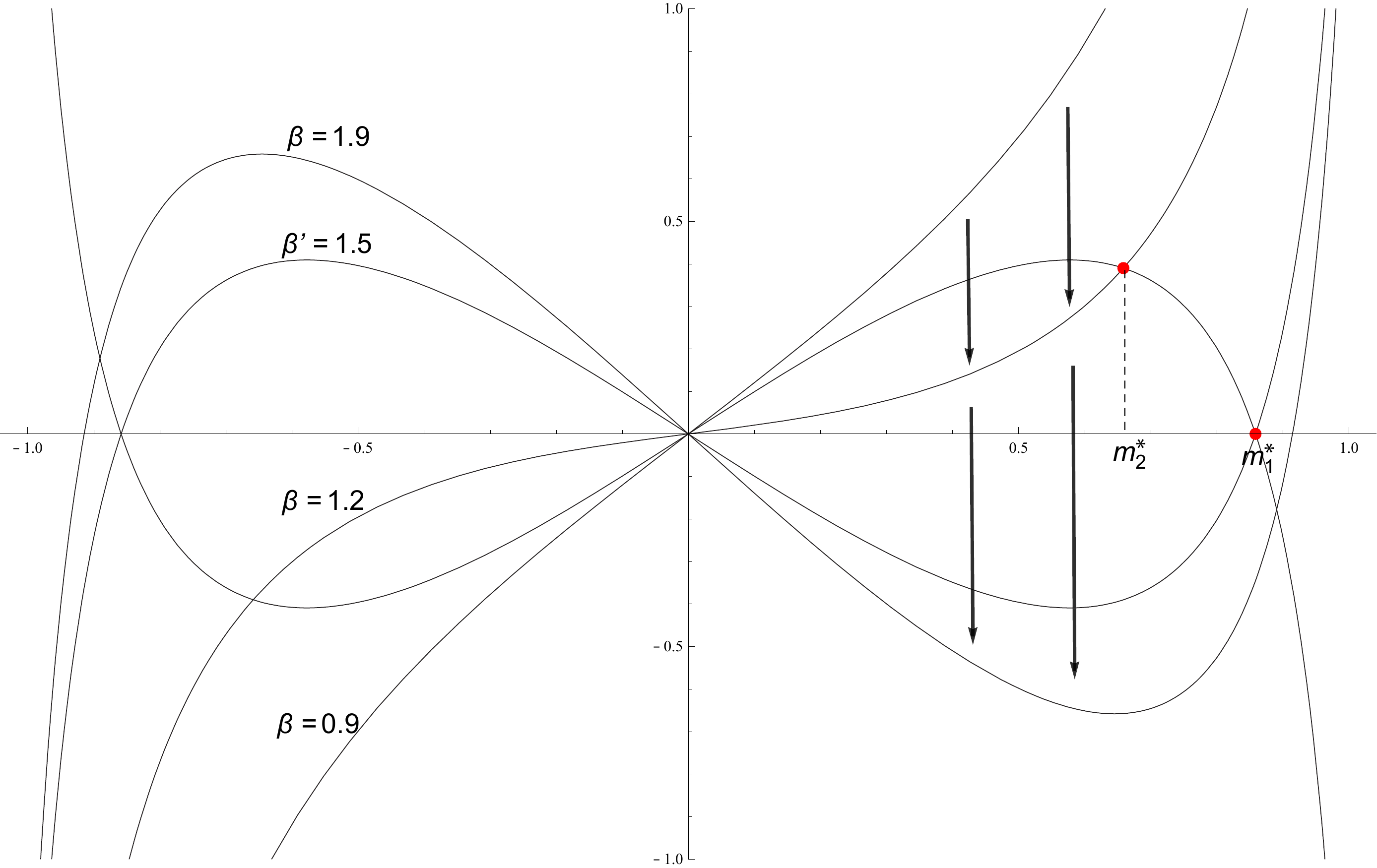}
\caption{Allowed-configurations curve for different $\b$ keeping $\b'$ constant}\label{fig:Intuitive}
\end{figure}

The allowed-configurations curve comes into the region of periodic motion 
and stays there when the following condition is satisfied
$$-f'(x)\Bigl |_{x=0}<g'(x)\Bigl |_{x=0}<f'(x)\Bigl |_{x=0},$$ which turns out to be just equivalent to 
\begin{equation}
\begin{split}
-(2\b'-2)<&2-4\b+2\b'<2\b'-2\cr
\end{split}
\end{equation}
or $1<\b<\b'$.
One can get an intuitive understanding of this mechanism from Figure \ref{fig:Intuitive}.
$\Cox$

%NUMERICS
\section{Numerical results: Typical paths, bad configurations, multiple histories, forbidden regions}

Since the variational problem with fixed endpoint \eqref{eq:EL} can not be solved in closed form 
unless the dynamics is independent, let us now describe some of the key features 
which are seen in numerical study. 
\begin{figure}[b]
\centering
\begin{tabular}{l l}
 \includegraphics[height=4.5cm]{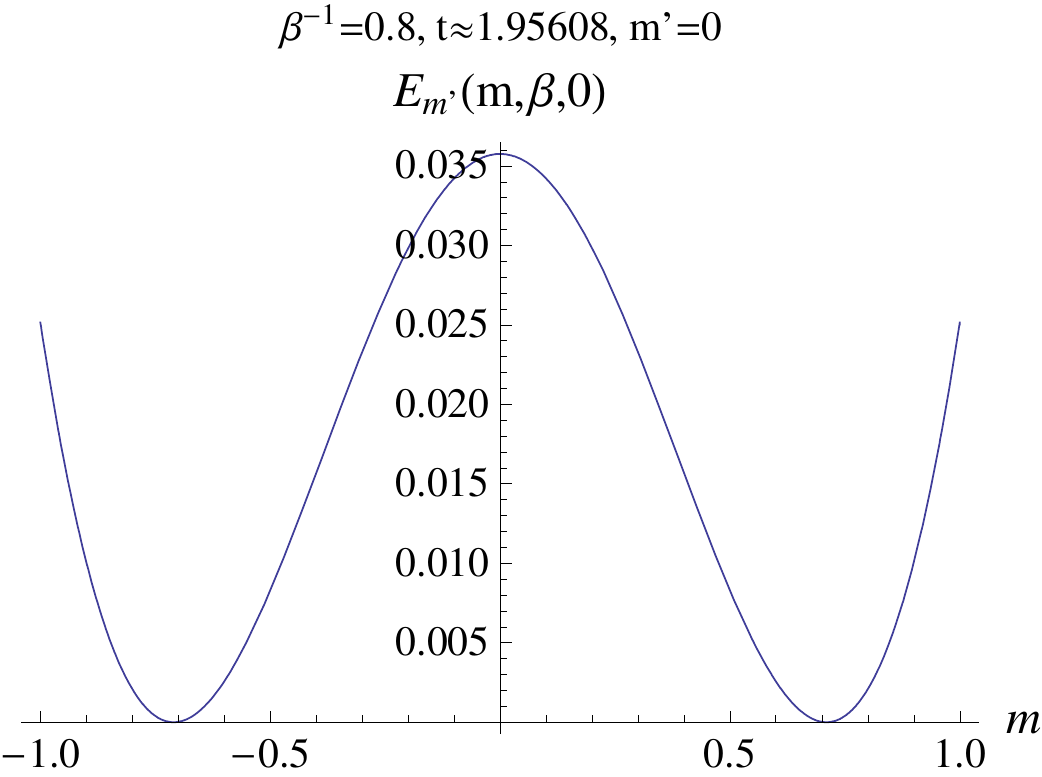}&
 \includegraphics[height=4.5cm]{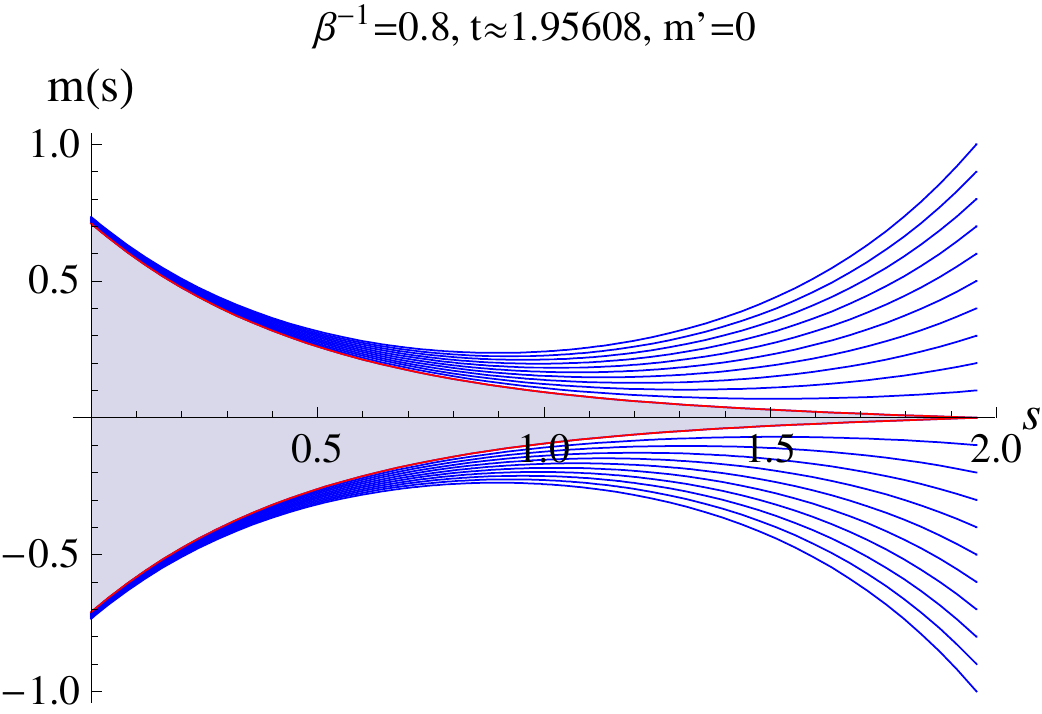}\\
  \small{Cost functional} & 
  \small{Corresponding history curves} \\
\end{tabular}
\caption{Symmetric forbidden regions}
\label{fig:sfb}
\end{figure}

For given conditioning $(\b',\b,t,m')$ a solution of
\eqref{eq:EL} with this set of parameters is called a {\em history curve}. 
Let us first discuss such curves for the example of the independent dynamics.
Figure \ref{fig:sfb} shows on the right such history curves conditioned to end at time $t$ 
at $m'$, for different values of $m'$. There is a jump in the optimal trajectory when we change 
$m'=0+$ to $m'=0-$. The associated cost functional at $m'=0$, depicted on the left, 
has two symmetric minima, and their minimizers are the two possible initial magnetization values. 
This is an example of a multiple history scenario.  
We call the regions showing on the right plot which cannot 
be visited by any integral curve {\em forbidden regions}. 

\begin{figure}[t]
\centering
\begin{tabular}{l l}
 \includegraphics[height=4.5cm]{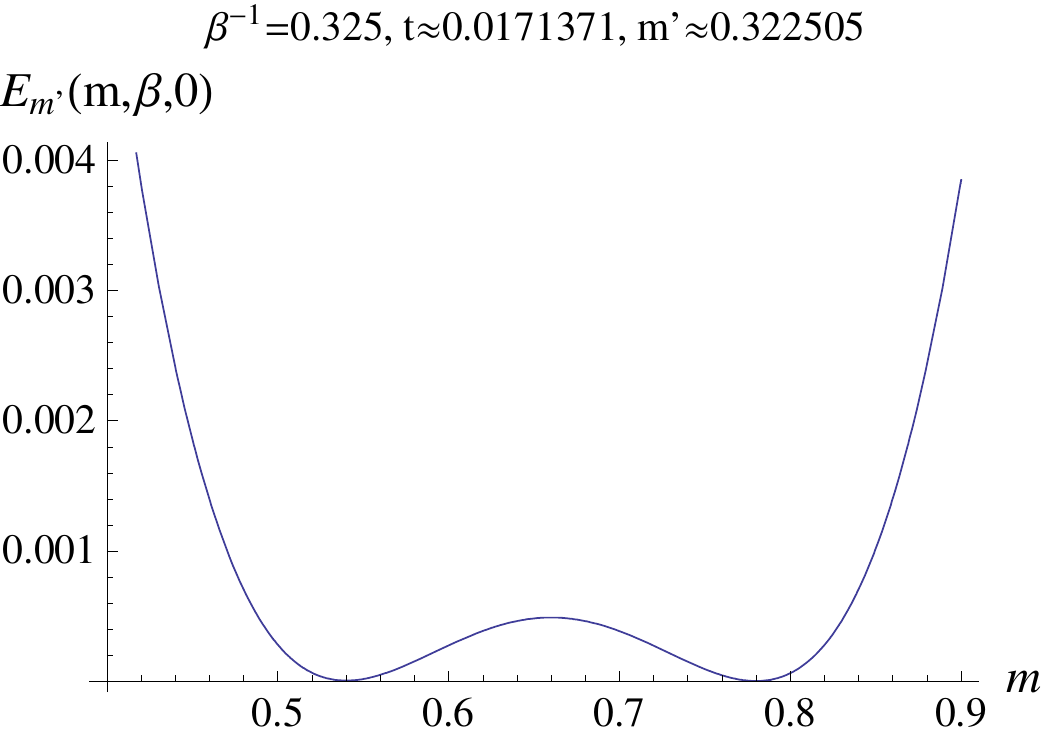}&
 \includegraphics[height=4.5cm]{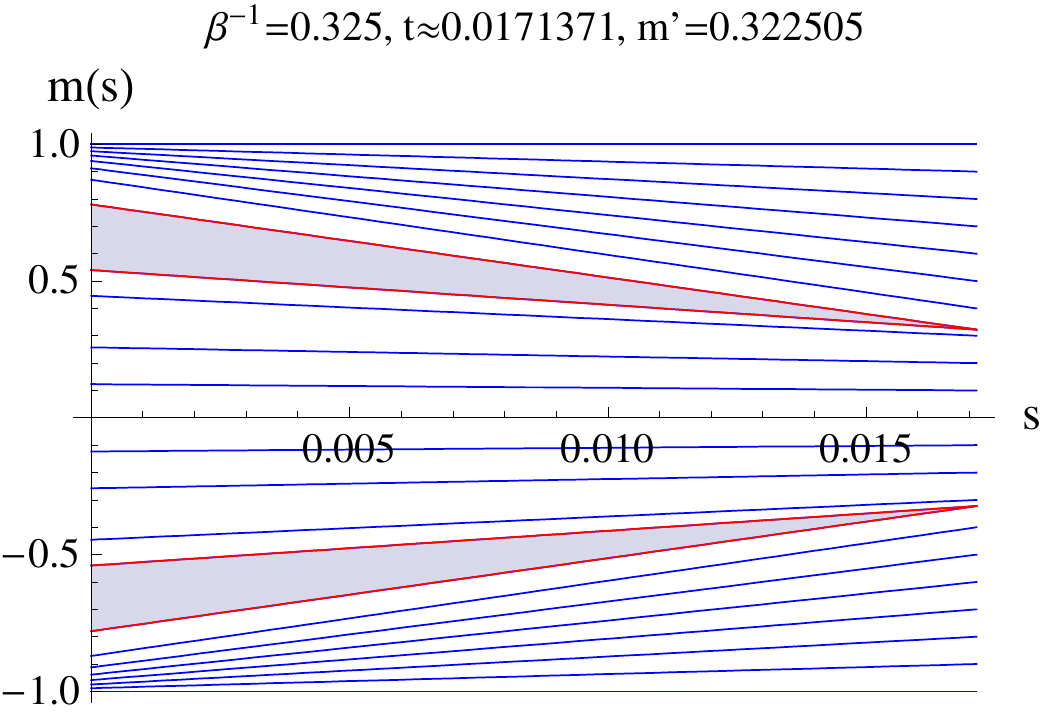}\\
  \small{Cost functional} & 
  \small{Corresponding history curves} \\
\end{tabular}
\caption{Non-symmetric forbidden region}
\label{fig:nsfb}
\end{figure}

Figure \ref{fig:nsfb} shows on the right history curves for the independent dynamics with 
a low initial temperature smaller than $\frac{2}{3}$ 
where symmetry-breaking in the set of bad configurations takes place. 
We see on the right two discontinuity points $m'$ and correspondingly two 
components of forbidden regions for the trajectory. The cost functional corresponding 
\begin{figure}[t]
\centering
\begin{tabular}{l l}
 \includegraphics[height=4.5cm]{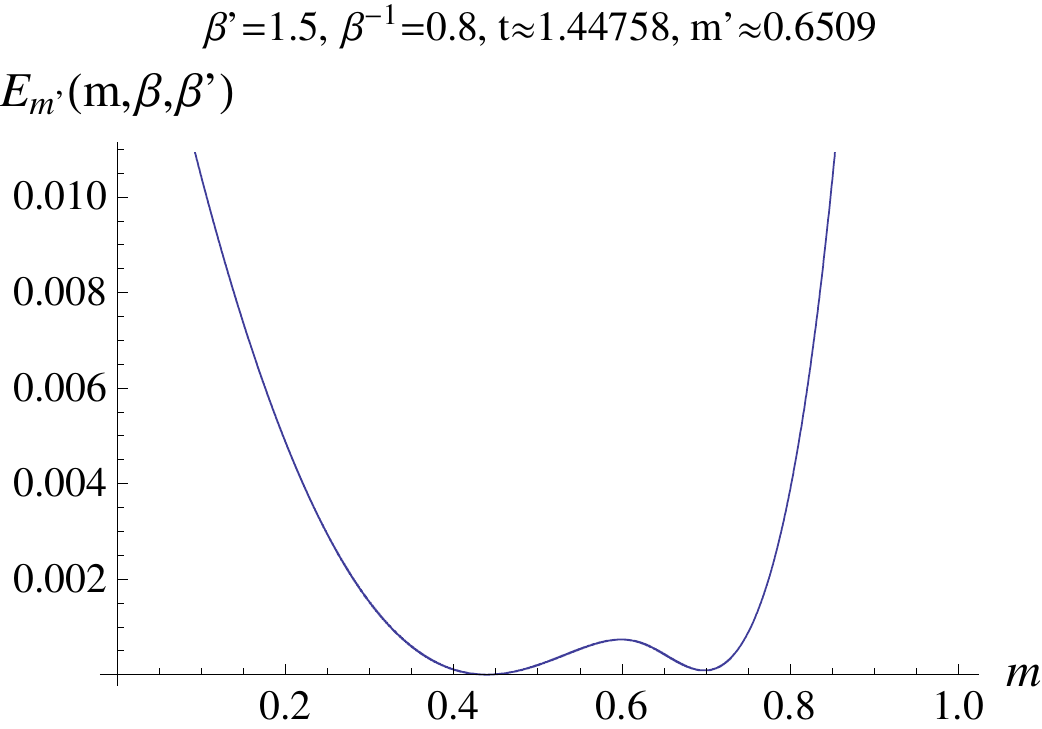}&
 \includegraphics[height=4.5cm]{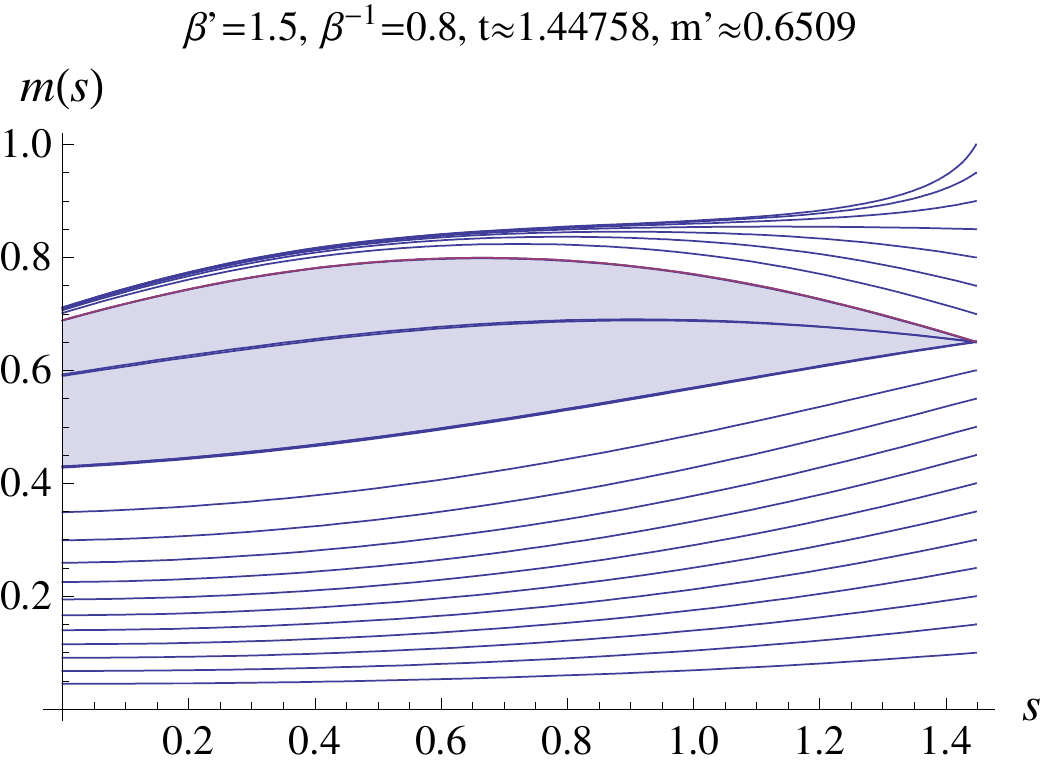}\\
  \small{Cost functional} & 
  \small{Corresponding history curves} \\
\end{tabular}
\caption{Forbidden region for $\b'= \frac32$}
\label{fig:sfb2}
\end{figure}

to the positive one of them is depicted on the right. 
Deformations of these pictures describe the phenomena for all temperatures of the dynamics, 
as long as the initial temperature is lower.

Finally, Figure \ref{fig:sfb2} displays history curves and 
cost functional at the critical conditioning for an example of cooling dynamics.

Next, let us fix $\b,\b'$ and describe the possible change of the set of bad configurations as a function of the time. 
Again we look at the independent dynamics first. 

\begin{figure}[t]
\centering
\begin{tabular}{l l}
	\subfloat[]{\label{fig:1m0mNG:A}\includegraphics[height=4.5cm]{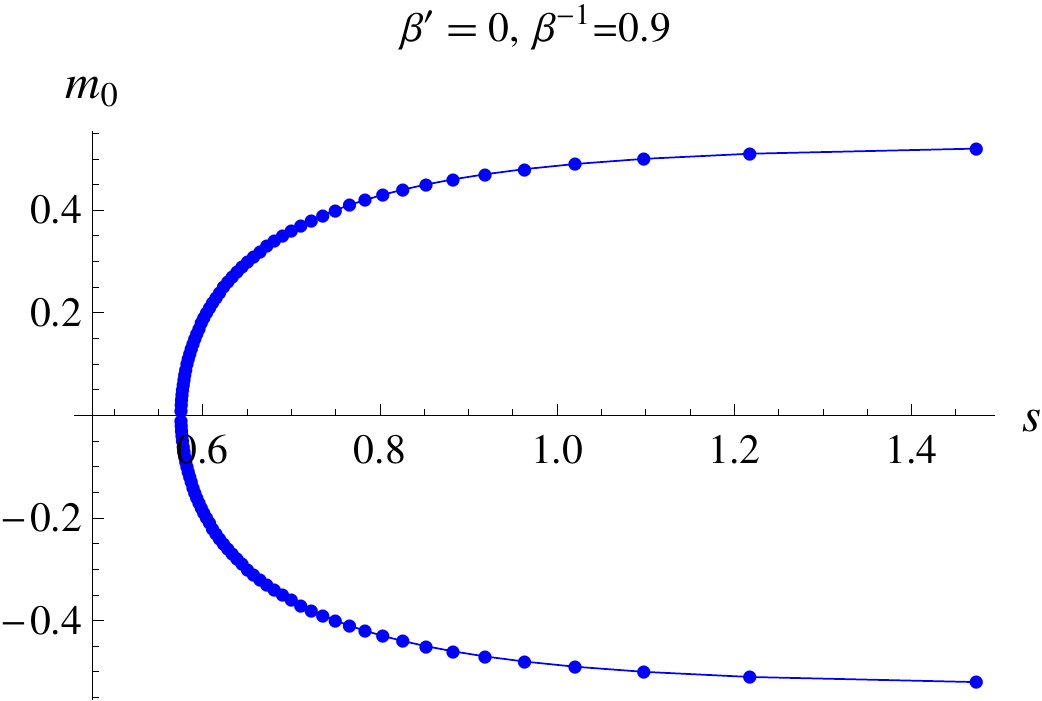}}&
	\subfloat[]{\label{fig:1m0mNG:B}\includegraphics[height=4.5cm]{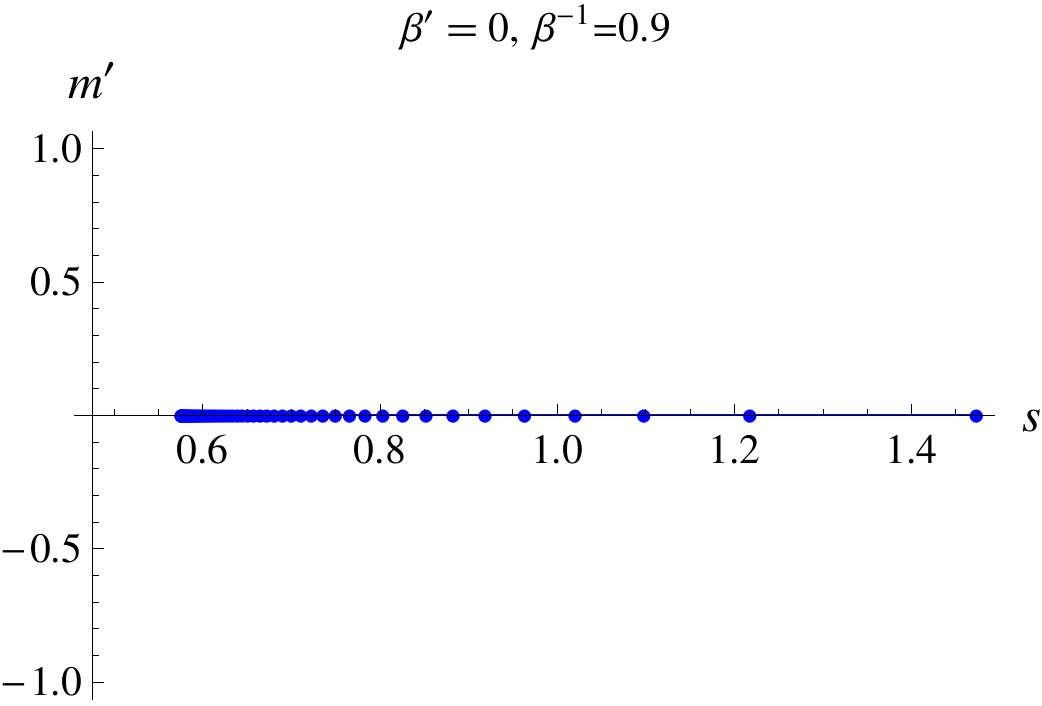}}\\
	
	\subfloat[]{\label{fig:1m0mNG:C}\includegraphics[height=4.5cm]{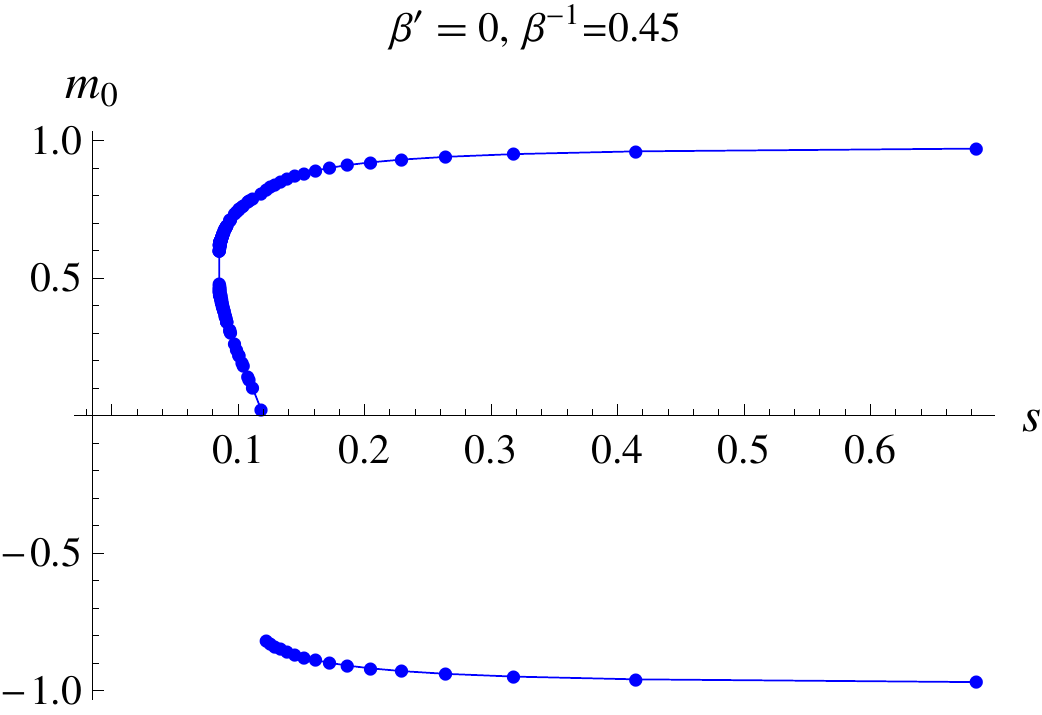}}&
	\subfloat[]{\label{fig:1m0mNG:D}\includegraphics[height=4.5cm]{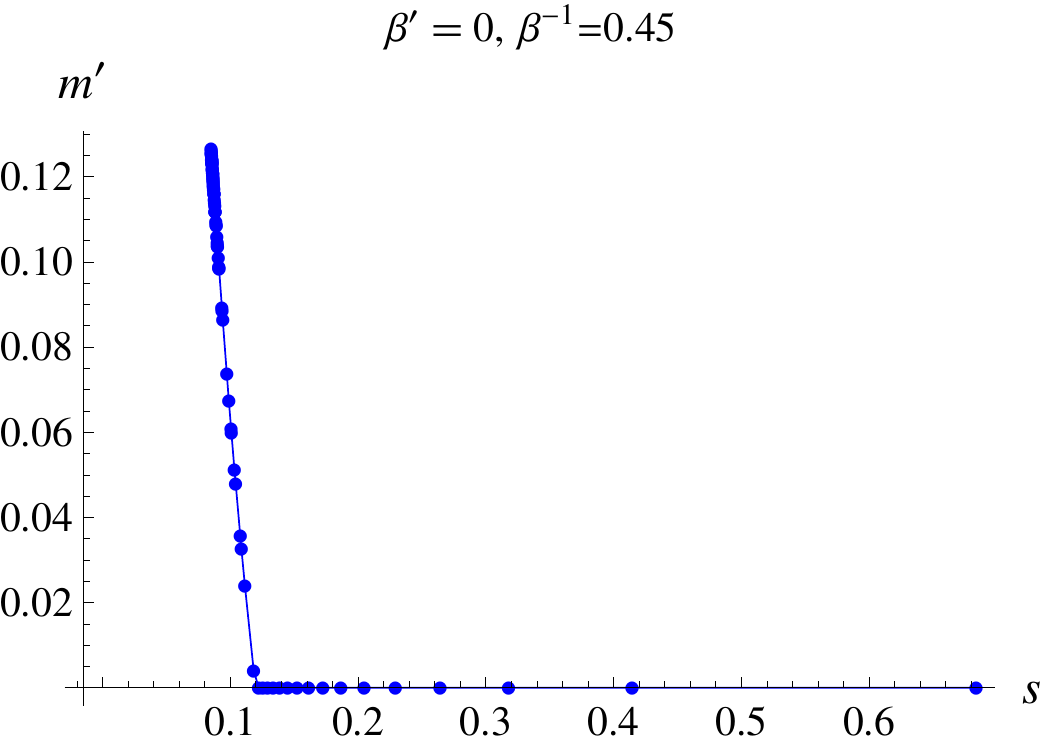}}\\
\end{tabular}
\caption{Bad configurations as function of time (right) and initial points of trajectories (left) $\b'=0$}\label{fig:1m0mNG}
\end{figure}

\begin{figure}[!htb]
\centering
\begin{tabular}{l l}
	\subfloat[]{\label{fig:3m0mNG:A}\includegraphics[height=4.5cm]{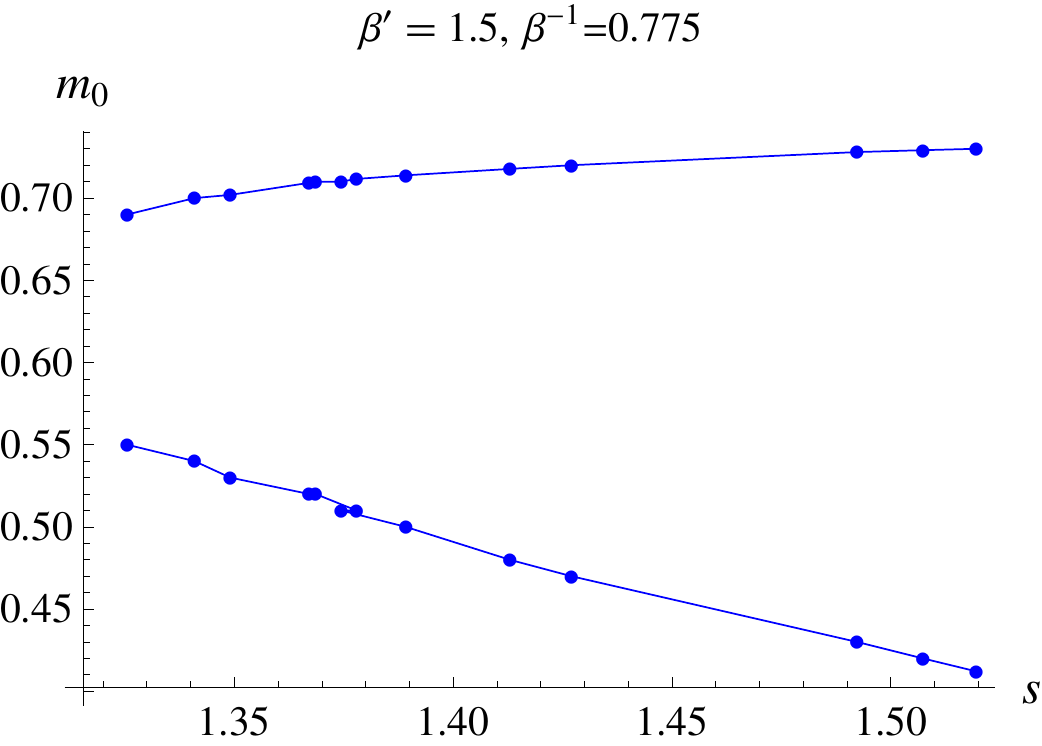}}&
	\subfloat[]{\label{fig:3m0mNG:B}\includegraphics[height=4.5cm]{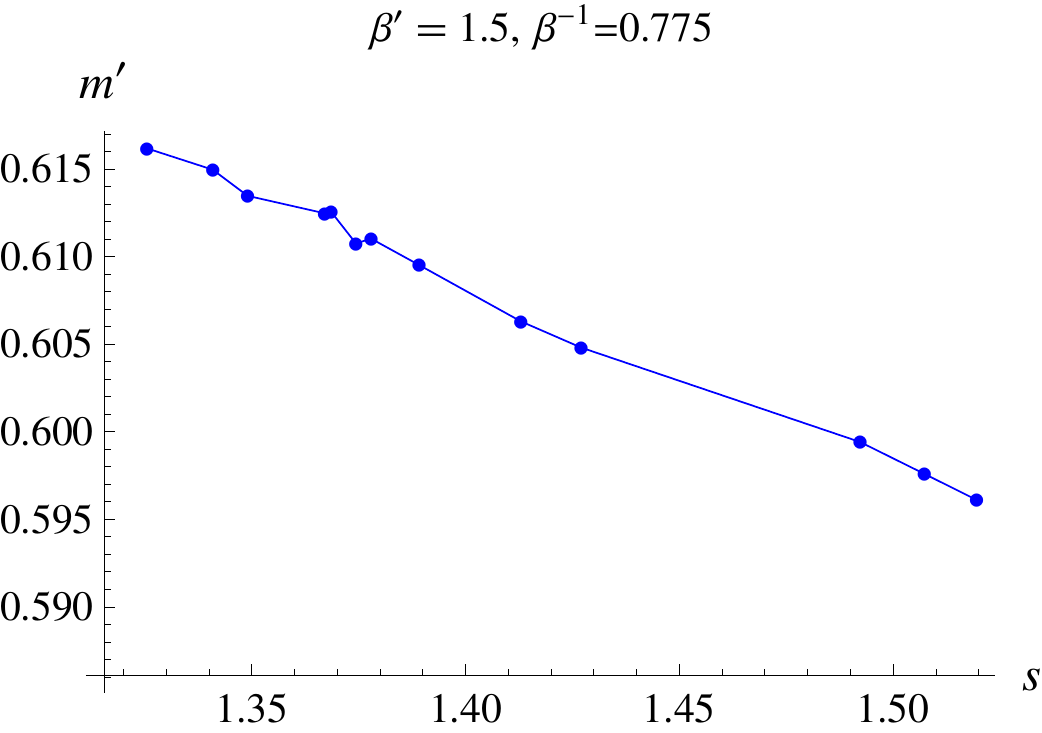}}\\
	
	\subfloat[]{\label{fig:3m0mNG:C}\includegraphics[height=4.5cm]{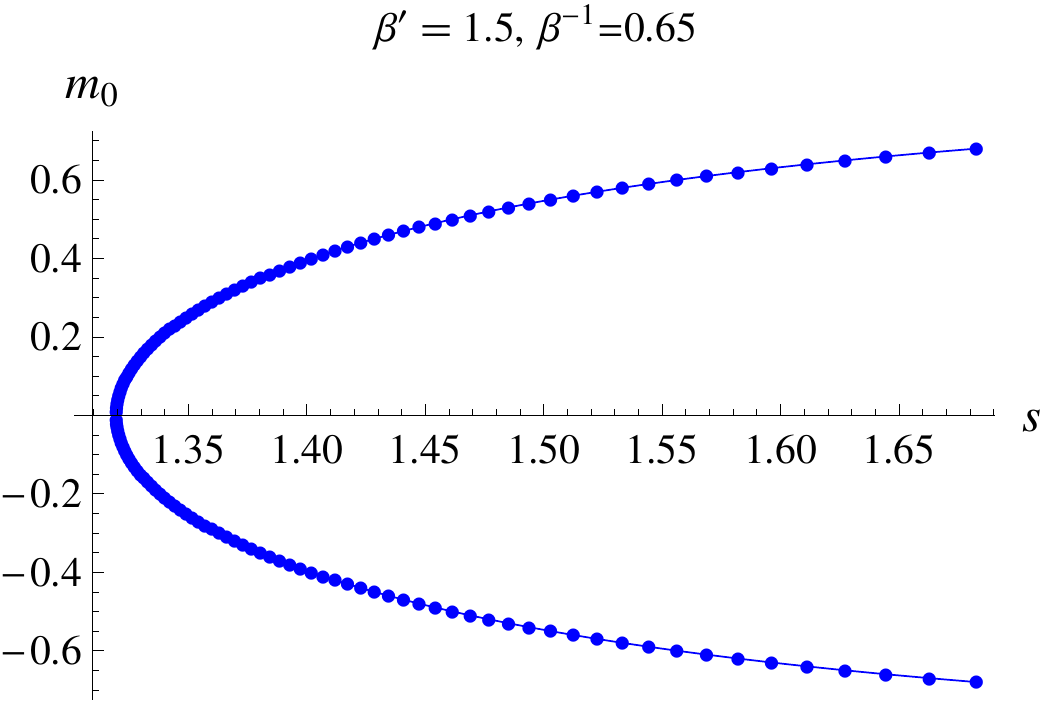}}&
	\subfloat[]{\label{fig:3m0mNG:D}\includegraphics[height=4.5cm]{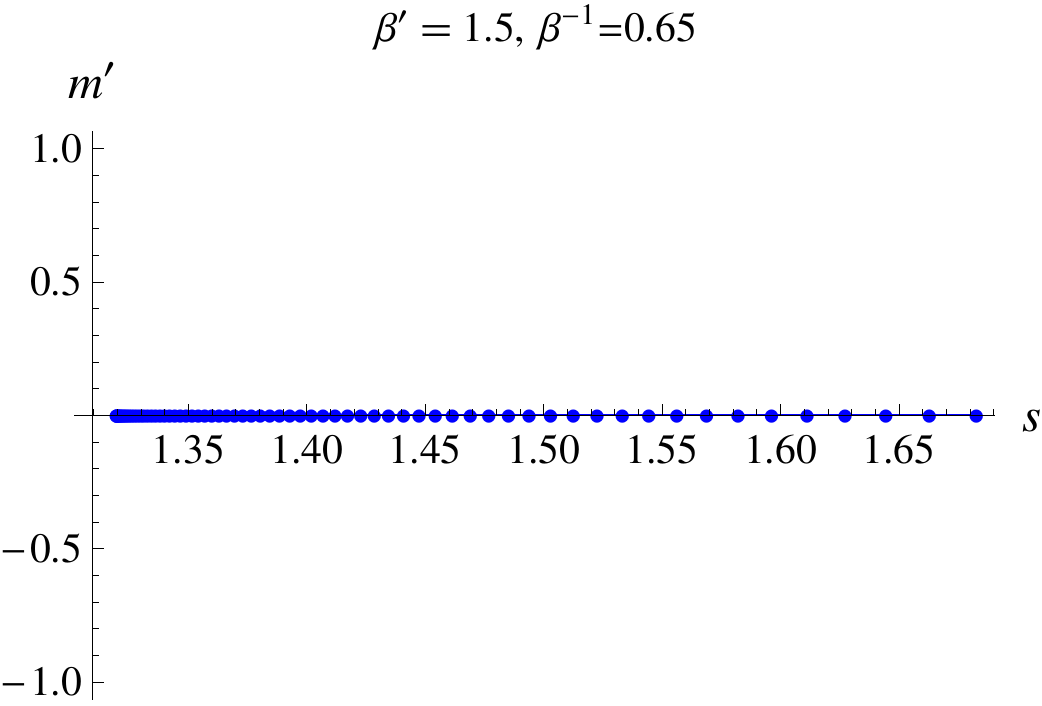}}\\
	
	\subfloat[]{\label{fig:3m0mNG:E}\includegraphics[height=4.5cm]{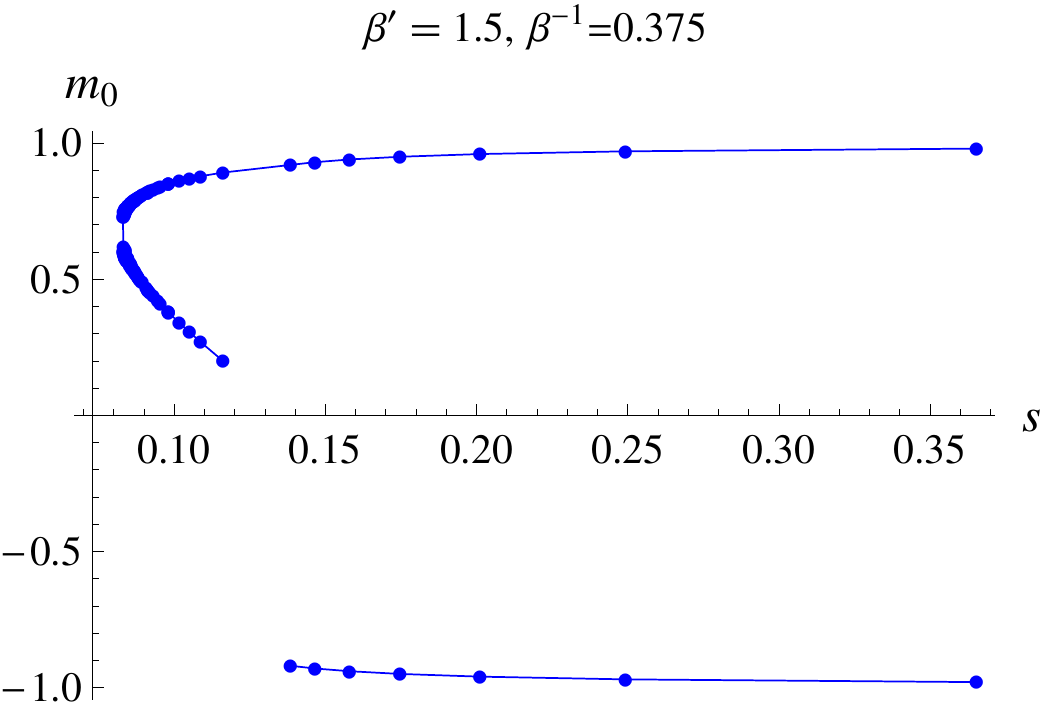}}&
	\subfloat[]{\label{fig:3m0mNG:F}\includegraphics[height=4.5cm]{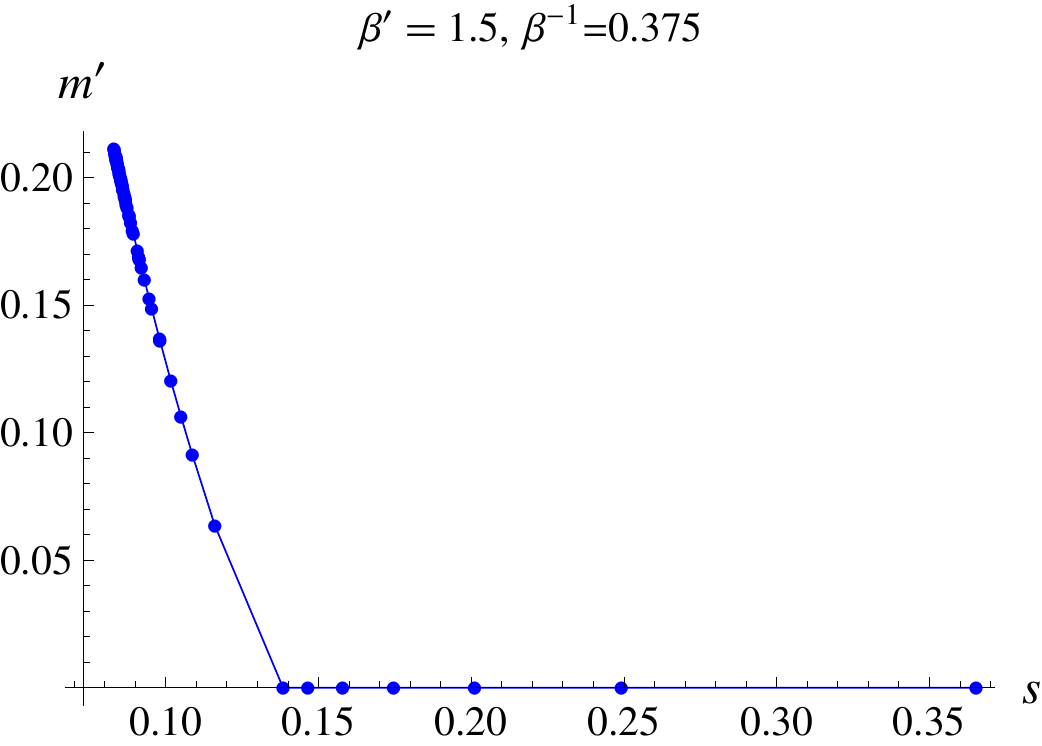}}\\
\end{tabular}
\caption{Bad configurations as function of time (right) and initial points of trajectories (left) - low-temperature dynamics $\b'=\frac32$}\label{fig:3m0mNG}
\end{figure}

The top line of Figure \ref{fig:1m0mNG} has an initial temperature in which non-Gibbsian behavior without symmetry-breaking takes place. In the picture \ref{fig:1m0mNG:B} we see the bad 
configurations $m'$ as a function of the time $s$ which were found numerically depicted by dots. 
Since $m'=0$ appears 
at a threshold time and stays to be the only bad configuration from that on, the graph 
of bad configurations is just a straight line starting at the threshold time. 
In the picture \ref{fig:1m0mNG:A} we see the corresponding initial 
points of the history curves which are conditioned to end at $m'$.  

The lower line of  \ref{fig:1m0mNG} has an initial temperature for which non-Gibbsian behavior with symmetry-breaking takes place,
in an intermediate time-interval. The right plot shows 
the corresponding non-negative branch of bad configurations $m'$. (By the symmetry of the model, 
taking the negative of these one obtains the full set of bad configurations.) The left plot shows 
the corresponding initial points of the history curves which are conditioned to end at the non-negative bad 
configurations $m'$ on the right. 

Finally, figure \ref{fig:3m0mNG} displays the time-evolution of bad configurations 
and their initial points for a low-temperature dynamics.  
The lowest line corresponds to heating from very low initial temperature and shows 
non-Gibbsianness with symmetry-breaking at an intermediate time-interval. 
The middle line corresponds to heating from an intermediate lower temperature 
and shows non-Gibbsianness without symmetry-breaking. These two mechanisms are 
known from high-temperature dynamics. Figures \ref{fig:3m0mNG:A} and \ref{fig:3m0mNG:B} correspond to cooling and 
shows data from the region of periodic orbits. 

Applying numerical integration of the Euler-Lagrange equations from initial 
conditions chosen on the allowed-configurations curve, check 
for intersecting trajectories and numerical computation of the cost function 
we can get (numerical approximations to) 
the array of bad quadruples $(\b,\b',t,m_{\text{pb}})$, augmented by the possible initial points. 
With this procedure we rederived the Gibbs-non-Gibbs phase diagram for $\b'=0$ (which 
was obtained earlier in\cite{KuLeNy}). 
Based on it we can draw the Gibbs-non-Gibbs phase-diagram at any dynamical temperature $\b'$. 
 An example for this was presented in the Introduction of the present paper in  figure 
 \ref{fig:gng} for a fixed low dynamical temperature. 

\newpage

\begin{section}{Appendix} 

%We must still give the proof of the unconstrained path large deviation principle. \bigskip 

\begin{subsection}{Sketch of proof of unconstrained path large deviation principle}

%{\bf Proof of Theorem \ref{thm:formJ}.} 
 Let us first consider first a simpler 
Markov jump process for the magnetization 
with transition rates which to not depend on the state $m$ of the process. The corresponding 
path large deviation principle can be built up as follows. The first ingredient 
is the large deviation principle for the Compound Poisson process. 

\begin{prop} Denote by $R_t$ a Poisson process with rate $1$. Denote by 
$\xi_i$, $i=1,2,\dots$ a sequence of i.i.d. random variables 
with exponential moment generating function $h(\l)=\E e^{\l \xi_1}$. 

Denote by $Z(t)=\sum_{i=1}^{R_t}\xi_i$ the associated Compound Poisson 
process. Define the rescaled paths by 
$Z_{N}(t)=\frac{1}{N}Z(N t)$.  

Define
\begin{equation}\begin{split}\label{eq:comppp}
J(v)= \sup_{\l}(v \l -  h(\l))+1
\end{split}
\end{equation} 
Then, at fixed $t$, as $N\uparrow \infty$, the distribution of the variable  $Z_{N}(t)$ satisfies a LDP in $\R$ with 
rate function $t J\left( \frac{v}{t}\right)$  and rate $N$. 
\end{prop}
This is a known theorem \cite{DemZei98} % (see \cite{Var02} for its history in relation to the history of large deviations),   
but it is instructive to see the proof for the sake of completeness.

{\bf Proof: }
Let us look at the logarithmic moment generating function $\Lambda_N(\l)$, defined by 
\begin{equation}\begin{split}
e^{\Lambda_N(\l)}:&=\E\Bigl(e^{Z_N(t)\l}\Bigr)
=\sum_{k=0}^\infty \E\Bigl(e^{\frac{\l}{N}\sum_{i=1}^k \xi_i}\Bigr)e^{-N t}\frac{(N t)^k}{k!}\cr
&=e^{\left(
-N t+ N t h\left(\frac{\l}{N}\right)
\right)}
\end{split}
\end{equation} 
Recall the G\"artner-Ellis theorem (Theorem 2.3.6. in \cite{DemZei98}, page 44) which states the following. 
Assume that  $\lim_N \frac{1}{N}\Lambda_N(N\l)=:\L(\l)$ exists. 
Then the distribution of the variable  $Z_{N}(t)$ satisfies a LDP on $\R$ with 
rate function $\L^*(v)$, which is the Fenchel-Legendre transform of $\L(\l)$. In our case we have equality even at finite $N$ of the form
$\L(\l)=\frac{1}{N}\Lambda_N(N\l)=t(h(\l)-1)$ and the Legendre transform gives 
\begin{equation}
\begin{split}
\L^*(v)&=\sup_{\l}(v \l-t(h(\l)-1))
=t J(\frac{v}{t})\cr
\end{split}
\end{equation} 
So the theorem follows. 
$\Cox$

In the next step we go from one-dimensional large deviations 
to large deviations of finite-imensional marginals in 
the path space of the Compound Poisson process.
This way of arguing corresponds to \cite{DemZei98} Lemma 5.1.8, page 178, which is a step
to prove Mogulskii's theorem. Recall that Mogulskii's theorem states that the paths 
of empirical averages of the form $\frac{1}{N}\sum_{i=1}^{\lfloor N t \rfloor} \xi_i$ satisfy 
a LDP. 

\begin{prop} For any decomposition $B=\{0<t_1 < t_2 <\dots < t_k \leq t\}$ of the time-interval 
$[0,t]$ and path $f:[0,t] \rightarrow \R$ denote by $\pi_B f$ the projection 
\begin{equation}\begin{split}
&\pi_B f= \left( f(t_1),f(t_2),\dots,f(t_k)\right)\cr
\end{split}
\end{equation}
Then the corresponding image measures $\pi_B (P_{N})=P_N\circ \pi^{-1}_B$ of 
the rescaled Compound Poisson process satisfy a large deviation principle in $\R^k$ 
with rate $N$ and rate function 
\begin{equation}\begin{split}
&J_B(z_1,\dots,z_k)=     t_1 J\Bigl( \frac{z_1}{t_1} \Bigr) 
+(t_2 -t_1) J\Bigl( \frac{z_2-z_1}{t_2-t_1} \Bigr) +\dots 
+(t_{k} -t_{k-1}) J\Bigl( \frac{z_k-z_{k-1}}{t_{k}-t_{k-1}} \Bigr)\cr
\end{split}
\end{equation}
where $J$ is defined in \eqref{eq:comppp}.
\end{prop}
The proof follows from putting together the result for the one-dimensional distributions as above. 
From here one gets the analogue of Mogulskii's theorem 
(see \cite{DemZei98} Theorem 5.1.2, page 176).  

\begin{thm} Denote by $P_{N}$ the law of the rescaled paths by 
$Z_{N}(s)=\frac{1}{N}Z(N s)$ of the Compound Poisson process as above, 
for $0\leq s \leq t$.  
  
Then the measures $P_{N}$ satisfy 
%in $L_{\infty}[0,t]$ 
a large deviation principle 
with rate $N$ and rate function given by the Lagrange functional  
$\LL(\phi)=\int_{0}^t J(\dot \phi(s))ds$  
where $J$ is defined in \eqref{eq:comppp}.
\end{thm}

We do not give a full proof here, but note that it  
follows by taking a supremum over the finite decompositions, and invoking 
the Dawson-G\"artner theorem, as explained in \cite{DemZei98}.

Up to this moment we have only treated the Compound Poisson process with
constant rates, which is not the case here, since we consider
state-dependent spin-flip dynamics, meaning state-dependent rates.
We need to justify rigorously that we can replace the contribution to the integral  
over the Lagrangian density for the infinitesimal time-interval $ds$ 
of the form $j(\dot \phi(s))ds$ by a term 
$j_{\b'}(\phi(s),\dot \phi(s))ds$ if the transition rates of the Markov chain depend on the state $\phi(s)$. 
In order to do that a comparison result is needed which compares a Markov 
chain with constant rates with the original Markov chain with state-dependent (but bounded) rates 
on the level of the logarithm of the exponential moment generating function, on small time-intervals.  
We are grateful to Frank Redig for communicating the following result to us which is an essential 
ingredient. Informally, the following lemma states that a jump process $z_N(t)$ on the discrete space with
constant jump-up, jump-down rates $c_\pm$ does not differ much from a process $m_N(t)$ with state-dependent rates $c_\pm(m_N(t))$
when the time interval $[0,\D t]$, where both processes are considered, is sufficiently small and $N$ is sufficiently large. This is due to the fact that the state of $m_N(\D t)$ cannot change a lot if $\D t$ is small and $N$ is large. At small times $m_N(t)$ can make not much more jumps than 
$N \D t$ jumps of a small height, therefore the state-dependent rates 
$c_\pm(m_N(0))$ and $c_\pm(m_N(\D t))$ will not vary much.

\begin{lem} {\bf F. Redig's useful lemma.}
Denote by $z_{N}(t)$ the Markov process on the discrete space 
$\{-1,-1+\frac{2}{N},\dots,1-\frac{2}{N},1\} $
%started at $m$ with constant non-zero rates $c_{\pm}(m)$ to go up of down by one step,  
started at $m_0$ with constant non-zero rates $c_{\pm}(m_0)$ to go up of down by one step,
%and by $m_N(t)$ the true process started at the same point $m$ with state-dependent rates
and by $m_N(t)$ the true process started at the same point $m_0$ with state-dependent rates  
$c_{\pm}(m)$ given by \eqref{eq:lin-genm}. 
Then 
\begin{equation}
\lim_{t\downarrow 0}\frac{1}{t}\sup_{m}\limsup_{N\uparrow\infty}
\frac{1}{N}\log \frac{\E \exp N \lambda z_N(t)}{\E_m \exp N \lambda m_N(t)}=0
\end{equation}
\end{lem}

The full proof will appear elsewhere, along with generalizations to more general local state spaces. 
Employing the lemma and going through suprema over finite partitions again, the proof 
of Theorem \ref{thm:formJ} is obtained where we still need to identify the form of the Lagrangian density. 
Let us start again with a Compound Poisson process with jumps of size $2$ with the distribution 
$P(\xi_1=2)=p, P(\xi_1=-2)=1-p$ where $p$ is fixed in the beginning.  
If we denote again $h(\l)=E e^{\l \xi_1}$  and  $J(v)= \sup_{\l}(v \l -  h(\l))+1$ we have that 
\begin{equation}\begin{split}
&J_p(v)=\frac{v}{2}\ln\Bigl(\frac{v+\sqrt{16 p-16 p^2+v^2}}{4 p} \Bigr)- \frac{1}{2} \sqrt{16 p-16 p^2+v^2} +1\cr
\end{split}
\end{equation}
as a solution of a quadratic equation shows. Choosing the rates in the rescaled Compound Poisson process 
to go up by $\frac{2}{N}$ (or down by $-\frac{2}{N}$) to match the rates in the generator $\hat L_{\b',N}$ we are led to choose 
\begin{equation}\label{bel}
\begin{split}
	p_{\b'}(m)
	=\frac{e^{2 \b' m} (1-m)}{e^{2 \b' m} (1-m)+(1+m)}\cr
\end{split}
\end{equation}
and this explains the form of the Lagrangian density  
$J_{\b'}(m,v)\equiv J_{p_{\b'}(m)}(v)$, after a small computation. 
This concludes our treatment of the proof. $\Cox$

\end{subsection}

\begin{subsection}{Free end-condition} To obtain the necessary condition \eqref{eq:ELf} for an extremum 
of the variational problem 
\begin{equation}\label{23x}
\begin{split}
%\phi\mapsto-\frac{\b \phi(0)^2}{2}+I(\phi(0))+J_{\b'}(\phi)
\phi\mapsto H(\phi(0))+I(\phi(0))+J_{\b'}(\phi)
\end{split}
\end{equation}
with $\phi(t)=m'$, use the standard procedure in calculus of 
variations adapted to 
the problem with a free left end: 
Consider a perturbation $\phi(s)+\e \D\phi(s)$ around the extremum 
$\phi(s)$, with a function $\D\phi(s)$ obeying the constaint $\D\phi(t)=0$ at the end-point 
but no constraint on $\D\phi(0)$ at the initial point.  
Plug $\phi(s)+\e \D\phi(s)$ into \eqref{23x}, expand to linear order in $\e$, and 
demand that the terms proportional to $\e$ vanish. 
Using partial integration under the $s$-integral one arrives at the Euler-Lagrange 
equation for $s$ in the interval between $0$ and $t$, and the additional 
free end condition at the initial point, the latter one following by demanding 
that the terms proportional to $\D\phi(0)$ have to vanish. 

Alternatively, the problem can be reformulated in terms 
of a problem with different Lagrange density but without 
initial punishment term, via incorporation of the initial term into the integrand. 
From here, we refer to \cite{Gelfand-Fomin} where free-end problems are discussed.  
\end{subsection}

\begin{subsection}{Hamiltonian and Lagrangian formalism} 
Following a suggestion of a referee, let us remark that, as an alternative to the derivation 
in terms of approximations by compound Poisson processes, 
the form of the Lagrangian can also be obtained 
going through the formalism of \cite{FeKu06}, see Chapter 1.4. Let us briefly sketch 
this procedure for the convenience of the reader. 
The jump process for 
the magnetization has a generator 
\begin{equation}\label{delabel}
\begin{split}
A_ N g(m)  &=N\Big( p_{\b'}(m)\bigl(g\bigl( 
m+\frac{2}{N}\bigr) -g(m)\bigr)+
(1-p_{\b'}(m))\bigl( g\bigl( 
m-\frac{2}{N}\bigr)  -g(m)\bigr)\Bigr)\cr
&=:N \int \Bigl( g(m+ \frac{z}{N})- g(m)\Bigr)\eta(m, dz) \cr
\end{split}
\end{equation}
 with $p_{\b'}(m)$ given by \eqref{bel}. This generator 
 is of the form treated in  \cite{FeKu06} with the obvious identification of $\eta(m, dz)$.  
From here one defines an operator $\hat\HH$ by the corresponding 
action on functions $f$ of the magnetization
of the form 
\begin{equation}
\begin{split}
(\hat\HH f)(m)&=  \int (e^{f'(m)z}-1)\eta(m, dz)\cr
&=p_{\b'}(m)(e^{2 f'(m)}-1)+ (1-p_{\b'}(m))(e^{- 2 f'(m)}-1)
\end{split}
\end{equation}
Following the formalism and replacing $f'(m)$ by a momentum variable $\l$ 
one defines the 
corresponding governing Hamiltonian function $\HH(m,\l)$ (or generalized energy) 
of the dynamics (not to be confused with the spin-Hamiltonian) 
\begin{equation} 
\begin{split}
&\HH(m,\l)=p_{\b'}(m)(e^{2 \l}-1)+ (1-p_{\b'}(m))(e^{- 2\l }-1)
\end{split}
\end{equation}
Performing a Legendre transform we arrive at the Lagrangian density 
$$j_{\b'}(m,\dot m)=\sup_{\l}\bigl(\l \dot m - \HH(m,\l)\bigr)$$ 
which governs the path large deviations (as stated in \cite{FeKu06}, page 12). 
This procedure was also employed for explicit computations in the infinite-temperature case 
in \cite{EnFerdenHRed2010}.  It is 
a computational exercise to verify that this approach reproduces the form 
of the Lagrangian density previously given. 
Note also that the integral  of motion we introduced in 
\eqref{eq:motionIntegral} is identical to the generalized energy $\HH(\phi(s),\dot \phi(s))$.

Let us remark that the study of the corresponding Hamiltonian-Jacobi equations 
is equivalent to the study of the Euler-Lagrange equations. 
\end{subsection}

\end{section}

\newpage

%\bibliographystyle{plain}
%\bibliography{biblio}

\end{document}